# Optical control of 4*f* orbital state in rare-earth metals

## Authors

Nele Thielemann-Kühn[1]*, Tim Amrhein[1], Wibke Bronsch[1,2], Somnath Jana[3]†, Niko Pontius[3], Robin Y. Engel[4]‡, Piter S. Miedema[4], Dominik Legut[5,6], Karel Carva[6], Unai Atxitia[1,8], Benjamin E. van Kuiken[7], Martin Teichmann[7], Robert E. Carley[7], Laurent Mercadier[7], Alexander Yaroslavtsev[7,9], Giuseppe Mercurio[7], Loïc Le Guyader[7], Naman Agarwal[7]§, Rafael Gort[7], Andres Scherz[7], Siarhei Dziarzhytski[4], Günter Brenner[4], Federico Pressacco[4], Ru-Pan Wang[4,11], Jan O. Schunck[4,11], Mangalika Sinha[4]¶, Martin Beye[4]‡, Gheorghe S. Chiuzbăian[10], Peter M. Oppeneer[9], Martin Weinelt[1]# and Christian Schüßler-Langeheine[3]#

## Affiliations

[1]Freie Universität Berlin, Fachbereich Physik, Arnimallee 14, 14195 Berlin, Germany.
[2]Elettra-Sincrotrone Trieste S.C.p.A., Strada statale 14 – km 163,5, 34149 Basovizza, Trieste, Italy.
[3]Helmholtz-Zentrum Berlin für Materialien und Energie GmbH, Albert-Einstein-Str. 15, 12489 Berlin, Germany.
[4]Deutsches Elektronen-Synchrotron DESY, Notkestraße 85, 22607 Hamburg, Germany.
[5]VSB - Technical University Ostrava, IT4Innovations, 708 00 Ostrava, Czech Republic.
[6]Charles University, Faculty of Mathematics and Physics, DCMP, 12116 Prague 2, Czech Republic.
[7]European XFEL, Holzkoppel 4, 22869 Schenefeld, Germany.
[8]Instituto de Ciencia de Materiales de Madrid, CSIC, Cantoblanco, 28049 Madrid, Spain.
[9]Uppsala University, Department of Physics and Astronomy, P.O. Box 516, 75120 Uppsala, Sweden.
[10]Sorbonne Université, CNRS, Laboratoire de Chimie Physique, - Matière et Rayonnement,75005 Paris, France.
[11]Universität Hamburg, Physics Department, Luruper Chaussee 149, 22761 Hamburg, Germany.

*Corresponding author. Email: nele.thielemann-kuehn@fu-berlin.de
† present address: Max Born Institute for Nonlinear Optics and Short Pulse Spectroscopy, 12489 Berlin, Germany.
‡ present address: Department of Physics, AlbaNova University Center, Stockholm University, S-10691 Stockholm, Sweden.
§ present address: Department of Physics and Astronomy (IFA), Aarhus University, Ny. Munkegade 120,
8000 Aarhus C, Denmark.
¶ present address: Institute for X-ray Physics, University of Göttingen, 37077, Göttingen, Germany.
# These authors contributed equally to this work.

**Abstract**
A change of orbital state alters the coupling between ions and their surroundings drastically. Orbital excitations are hence key to understand and control interaction of ions. Rare-earth (RE) elements with strong magneto-crystalline anisotropy (MCA) are important ingredients for magnetic devices. Thus, control of their localized 4*f* magnetic moments and anisotropy is one major challenge in ultrafast spin physics. With time-resolved X-ray absorption and resonant inelastic scattering experiments, we show for Tb metal that 4*f*-electronic excitations out of the ground state multiplet occur after optical pumping. These excitations are driven by inelastic 5*d*-4*f*-electron scattering, alter the 4*f*-orbital state and consequently the MCA with important implications for magnetization dynamics in 4*f*-metals, and more general for the excitation of localized electronic states in correlated materials.

## Introduction

In ultrafast spin dynamics in metals, optically induced transitions in valence electronic states are the first excitation step. Recent experiments and theory identified direct consequences of these initial excitations for the spin dynamics: optical inter-site spin transfer (OISTR) between different atomic sites of transition metal (TM) compounds leads to enhanced speed and efficiency of demagnetization (*1*). Excitations of 4*f* electronic states in RE metals have not been considered to contribute to magnetization dynamics so far. X-ray magnetic circular dichroism studies, searching for separate spin and orbital dynamics in TM-RE ferrimagnets (*2,3*) were implicitly looking for changes of the *f* electronic state, albeit without stating that. They ultimately remained without conclusions. In contrast to valence states in TMs, which can be easily optically excited, the dipole selection rules make the cross section for optical 4*f* excitations in RE metals very small (*4*) in particular for pump-photon energies not resonant with 4*f*-4*f* transitions. Therefore, laser-driven 4*f*-spin dynamics must result from coupling to other degrees of freedom, such as to the lattice via phonons or to the directly laser excited 5*d*6*s* valence electrons. While non-parallel 4*f*-spin order can be efficiently quenched via angular momentum redistribution (*5,6*), the speed of demagnetization in 4*f*-ferromagnets is closely connected to the 4*f*-spin-lattice coupling (*7-15*). Launching phonons in 4*f* metals requires energies of few meV, which is enough to generate spin waves (*16*) and excitations within the 4*f* ground state multiplet (*17*), but not enough to reach higher lying 4*f* electronic multiplet states: the lowest-lying *f*-*f* multiplet transitions that change the total angular momentum *J* usually require energies of hundreds of meV (*18*).

The spin-polarized itinerant 5*d* electrons mediate the indirect exchange coupling between the 4*f* states of neighboring atoms (Ruderman-Kittel-Kasuya-Yosida (RKKY) interaction) and thus enable in equilibrium long-range order of the strongly localized 4*f* magnetic moments (*16).* Distinct 5*d* and 4*f* magnetic order dynamics observed in state selective studies on different RE metals (*6,7,9*) indicate a more intricate interplay of 4*f* and 5*d* states in non-equilibrium, but details of the 5*d*-4*f* interaction after optical pumping have rarely been explored. Partly, this is the case because a combined theoretical description of *d* and *f* electronic structure is difficult (*19*); magnetic dynamics simulation based on models like Landau-Lifshitz-Gilbert equation (*7,20-22*) or the Microscopic Three Temperature Model (*14,23,24*) usually treat the 4*f* and 5*d* magnetic systems as a single spin system. We present here an analysis of the 5*d*-4*f* coupling mechanisms in non-equilibrium and show that this coupling induces changes in the 4*f* electronic state that are directly linked to spin and orbital dynamics: orbital state transitions altering the total angular momentum *J,* have immediate consequences for the 4*f*-spin lattice coupling, which should be reflected in spin-dynamics.

With new pulsed X-ray sources and instrumentation at the European X-ray Free Electron Laser (EuXFEL) and the Free Electron Laser in Hamburg (FLASH), experimental approaches have recently become feasible that merge high energy and time resolution with core-state selective probing. For detecting the 4*f* electronic dynamics we combined two complementary methods: X-ray absorption spectroscopy (XAS) and resonant inelastic X-ray scattering (RIXS), as schematically illustrated in Fig. 1A. The time-resolved XAS and RIXS experiments were performed at EuXFEL and FLASH, respectively (see Figs. 1B and 1C for experimental schemes). XAS (at the 3*d*-4*f* transition) probes the full 4*f* multiplet and hence the sum of all changes to the electronic structure (*25*). RIXS (at the 4*d*-4*f* transition) allows for a selective detection of a specific electronic excitation with high sensitivity (*26*). We combined these two techniques to track femtosecond changes in the 4*f* multiplet structure after exciting the 5*d*6*s* valence electrons in paramagnetic hcp Tb metal with near-infrared laser pulses, and we observed distinct 4*f* orbital excitations. We show that variations of the 4*f* state are primarily driven by 5*d*-4*f* electron-electron scattering, but we also find indications of 5*d*-4*f* electron transfer in the 4*f* electron dynamics. The effect is not small: Under our experimental conditions we observe 5*d*-4*f* electronic scattering to excite about 20 % of the atoms in the probed volume. And it is relevant: The altered spin and orbital angular momenta of the excited 4*f* electronic state affect the coupling of the 4*f* magnetic system to the environment: Density functional theory (DFT) calculations show that the observed orbital transitions can affect the MCA such that the magnetic easy axis locally flips on a femtosecond time scale.

## Results

<u>Time-resolved XAS and RIXS</u>

As depicted by the experimental scheme in Fig. 1B, we recorded XAS spectra by probing the transmission through a thin Tb layer deposited on a silicon nitride membrane. We excited our sample with 800-nm-laser pulses (1.55 eV) and probed the XAS signal with monochromatized X-ray pulses from EuXFEL (see Materials and Methods). Time and energy resolution in the experiment was 65 fs and 350 meV, respectively. Figure 2A shows spectra recorded at the $M_5$ resonance, a transition into $3d^9 4f^9$ states, which leaves behind a $3d_{5/2}$ core-hole. The blue dots reflect the Tb $M_5$ absorption spectrum with 4*f* electrons in the $4f^8\ {}^7F_6$ ground state ('unpumped'). Recorded 150 fs after near-IR-laser excitation the orange dots ('pumped') show a spectrum with varied line shape, indicating that pumping has changed the 4*f* electronic state. A reference measurement on Gd under identical conditions (see Supplementary Materials (SM), Fig. S1), shows no pump effect in the XAS signal. As we explain further below this confirms that the effects seen for Tb XAS are in fact caused by changes of the 4*f* electronic structure.

In the RIXS setup shown in Fig. 1C, we excited a Tb metal layer epitaxially grown on a W(110) crystal with 1030 nm laser pulses (1.2 eV) and recorded the RIXS signal with monochromatic X-ray pulses from the FLASH (see Materials and Methods). The time resolution was 300 fs and we recorded spectra with an energy resolution of 140 meV. The RIXS spectra measured at the $N_{4,5}$ ($4d \rightarrow 4f$) transition (147.2 eV) are shown in Fig. 2B; the peaks in the spectra are related to electronic transitions into different final states in the 4*f* shell (Fig. 1A). The orange dots represent the spectrum 300 fs after excitation with the pump-laser pulse ('pumped'), which differs from that of the sample in the $4f^8\ {}^7F_6$ ground state (blue dots, 'unpumped'). Both spectra are plotted over energy loss, i.e., the energy difference between incoming and scattered photons. The spectrum of the pumped sample shows a transfer of intensity from the main features at 2.7 and 3.4 eV to losses of around 2.4 and 3.1 eV, respectively, again indicating an excited 4*f* initial state.

In order to identify the particular excitations in the 4*f* multiplet we describe the 4*f* ground and excited state XAS and RIXS spectra with atomic multiplet calculations (*27,28*) (see SM for details). Based on these calculated spectra we simulate the spectra for the pumped and unpumped sample (see Figs. 2C & 2D), as well as the differential change (relative change compared to the ground state spectrum in %) in XAS and RIXS, respectively (see. Figs. 2E & 2F).

4*f*-5*d* inelastic scattering

We start with analyzing the RIXS spectra presented in Figs. 2B. The main pump-induced change is a shift of the two ground-state RIXS features at around 2.64 eV and 3.4 eV by about 0.26 eV to lower energies (black horizontal arrows in Fig. 2D). The observed energy difference matches the lowest-lying multiplet excitation that involves a change of $J$ in the Tb $4f^8$ multiplets ($^7F_6 \rightarrow {}^7F_5$, $\Delta J = -1$). We verify this interpretation by the simulation shown in Fig. 2D and Fig. 2F. We can reproduce the pump-induced spectral changes by including an admixture of 4*f*-excited-states to the description of the ground-state spectrum. As depicted by the orange line in Fig. 2D and the green line in Fig. 2F, adding a contribution of about 5% $^7F_5$ to the $^7F_6$ ground state spectrum (blue line in Fig. 2D) yields already a reasonably good description of the transient changes in the RIXS data.

The RIXS cross-section for specific 4*f* multiplets across the Tb $N_{4,5}$ resonance strongly depends on the incoming photon energy; their spectral weight is therefore not a good measure of the admixture of excited states in the probed volume. We better quantify the overall contribution of $^7F_5$ excitations from the XAS spectra measured at 150 fs pump-probe delay (Fig. 2A). The purple solid line in Fig. 2E with an admixture of about (16 ± 1) % $^7F_5$ qualitatively describes the overall shape of the differential absorption, thus giving a measure of the total contribution of $^7F_5$ excited states at 150 fs delay.

How is this excitation possible? As argued above, 4*f*-4*f* transitions directly induced by the pump pulse are negligible. For optical *d-f* excitations from the itinerant 5*d* valence states into the localized 4*f* states or vice versa our pump-laser-photon energies are too low: Such excitations require energies of 2.8 and 2.3 eV, respectively (*29*), while we pump with 1.55 eV photons for XAS and 1.2 eV photons for RIXS (two-photon absorption processes are negligible under our experimental conditions; see SM). The pump-laser excitation hence cannot alter the 4*f* state directly, but it heats the valence electrons of mixed 5*d* and 6*s* character, which then can transfer the excitation to the 4*f* system.

That such a 5*d*6*s*-4*f* energy transfer is the relevant process here, is revealed by the temporal evolution of XAS and RIXS spectral changes. In Fig. 3 we present the differential absorption measured at 1236 eV (marked by the vertical line in Figs. 2A and 2E) versus the delay between the IR-pump and X-ray-probe pulse (black data points). The absorption at this energy drops rapidly with a decay constant of 70 fs, reaches a maximum deviation from the ground state of 3 % after 200 fs and recovers with a time constant of about 3 ps (see SM, Fig. S2). The transient change in the energy-loss region 2.7 – 3.2 eV of the RIXS spectra (hatched area in Figs. 2B and 2F) recorded with a lower time resolution of 300 fs follows the same trend (see Fig. 3, green markers). These observed time scales resemble those for optically excited valence electrons in metals (*30,31*): The blue line in Fig. 3 shows the temperature of the 5*d*6*s* electronic system calculated with a two-temperature model (2TM), involving 5*d*6*s*-electrons and phonons (see SM). The electron-temperature curve qualitatively matches the transient differential change in the XAS and RIXS signal at 1236 eV. We infer that the 4*f* multiplet is only excited, in the presence of *hot* 5*d*6*s* electrons. We assign the mechanism for the $^7F_6 \rightarrow$ $^7F_5$ excitations in the 4*f* shell to inelastic 5*d*-4*f* electronic scattering, involving transfer of energy and angular momentum from the optically excited 5*d*6*s* electrons to the 4*f* system. As only *5d6s* electrons with enough energy can induce the 4*f* transition, we better compare the differential spectral changes with the transient population of those hot electrons, that actually can transfer the energy required for the $^7F_6 \rightarrow {}^7F_5$ transition by filling a 5*d* holes at respectively lower energies (black line in Fig.3, see SM for details). This curve yields an even better fit to the XAS and RIXS data, supporting our assumption about the underlying scattering process.

In Fig. 4 we illustrate the supposed mechanism in a total-energy scheme for the relevant channels in XAS and RIXS, explaining the spectroscopic signatures of the 4*f*-multiplet excitation. In XAS at the $M_5$ resonance 3*d* electrons are excited from the initial state $3d^{10}4d^{10}4f^8(5d6s)^3$ to the $3d^94f^9$ final states (purple arrows). In RIXS at the $N_4$ edge, 4*d* electrons are excited from the initial state $3d^{10}4d^{10}4f^8(5d6s)^3$ to the $4d^94f^9$ intermediate states (blue arrows). In the Tb ground state, the $4f^8$ electrons occupy the $^7F_6$ multiplet. Optical excitation by the pump laser elevates the complete ground state (red arrow) by exciting the $(5d6s)^3$ electrons. They rapidly thermalize by screened Coulomb interaction and become Fermi distributed (indicated by the red area in the middle section of Fig. 4). By inelastic electronic 5*d*-4*f* scattering (Feynman diagram in the top middle section) energy and angular momentum is transferred between the $(5d6s)^3$ and $4f^8$ electronic system. A 4*f* electron is excited from $^7F_6$ to $^7F_5$ and a 5*d* electron fills a hole of lower energy. On sites where this happens, the initial state in XAS and RIXS changes (right side of Fig. 4). In XAS this affects the spectral line shape (compare Fig. 2A). In RIXS the raise of the initial state energy causes a red shift of the energy loss features (compare Figs. 2B & 2D with the energy loss values given in the right section of Fig. 4). The $^7F_6 \rightarrow {}^7F_5$ transition implies a change of angular momentum $\Delta J = -1$. A spin flip, altering the 5*d* spin by $\Delta S = 1$ would conserve the total angular momentum in the scattering process.

How about the excitations of energetically higher lying multiplets $^7F_J$ with $J = 4,3,2,1,0$? Such transitions require an angular momentum transfer $\Delta J < -1$. Either multiple 5*d*-4*f* electronic scattering events each involving $\Delta S = 1$ or a transfer of angular momentum from the reservoir of quenched 5*d* orbital momenta might permit higher $\Delta J$ changes. Indeed, by including the contribution of higher $4f^8$ multiplets to the 'pumped' XAS spectrum (assuming a thermal occupation, weighted with a Boltzmann factor and degeneracy $2J + 1$; see SM for details) improves the fit to the differential XAS (dashed purple line in Fig. 2E). Also, the asymmetric peak profile in the differential RIXS (black markers in Fig. 2F) could come from energetically higher lying excitations within the 4*f* states. Yet, with the presently available experimental data, we cannot answer the question unequivocally: The slightly better fit to the XAS data does not really prove these transitions; and the discontinues variation of energy resolution along the energy axis in the RIXS spectra makes the detection of the corresponding RIXS features difficult. We further found no evidence for these higher $4f^8$ states from features on the anti-Stokes side of the spectra (see SM for details). But we find indications for other 4*f* excitations as we discuss in the following.

5*d*-4*f* electron transfer

We notice that the shape of the pumped XAS spectrum is not constant but changes with pump-probe delay. Figure 5A presents an energy vs. delay map illustrating the differential absorption up to 430 fs after pumping. Difference spectra (dots in Figs. 5B & 5C) averaged over two successive delay ranges (I) and (II) (marked in Fig. 5A by black vertical lines and color-coded in Fig. 3) show that the dip at 1235.8 eV becomes more pronounced at larger pump-probe delays while the one at 1237.5 eV changes only slightly: Obviously the 4*f* state evolves in time. The simulated differential XAS changes from excitation of higher $^7F_J$ multiplets ($J$ = 4,3,2,1,0) do not explain the evolution of the pronounced pump effect at 1235.8 eV (see SM for spectral contribution of different $^7F_J$ multiplets). Therefore, we discuss the possibility of excitation into $4f^9$ and $4f^7$ multiplets.

The pump-excitation energy in our experiment is not sufficient to directly excite states with varied 4*f* electron number: The transitions to a $4f^9$ ($^6H_{15/2}$) or $4f^7$ ($^8S_{7/2}$) multiplet requires 2.8 eV and 2.3 eV, respectively (*29*). However, once thermalized, the Fermi distribution of the hot valence electrons extends beyond the initial pump energy; electrons at 2.8 eV above the Fermi energy can fill an empty 4*f* state and holes 2.3 eV below $E_F$ can accept the most lightly bound minority spin electron of Tb via elastic tunnelling processes (see illustration in Fig. 5D).

In fact, by including $4f^9$ and $4f^7$ multiplets in our XAS simulation we can achieve a quantitative description of the reported spectral changes at 150 fs delay in Fig. 2A. The best fit (blue solid line in Fig. 2E) contains (20 ± 1) % of the $^7F_5$ excited $4f^8$ states, and (4.5 ± 0.7) % of $4f^9$ and (3.3 ± 0.2) % of $4f^7$ electron-transfer contributions. We find our assumption of $4f^7$ and $4f^9$ contributions supported by the spectral changes on short delay scales presented in Fig. 5. As these states become populated only after the 5*d*6*s*-electron system has thermalized, we expect lower contributions before the maximum electron temperature is reached. In fact, we can describe the spectral changes with delay in Figs. 5B and 5C by different $4f^7$ and $4f^9$ contributions (solid lines). The $4f^9$ weight grows from (1.2 ± 0.5) % in interval (I) to (4.8 ± 0.4) % in interval (II), for $4f^7$ it grows from (2 ± 0.2) % to (2.8 ± 0.2) %.

Including $4f^7$ and $4f^9$ multiplets in the simulation requires some assumptions. The spectral shape itself can readily be simulated (and experimentally observed in Gd and Dy, respectively). Unlike for the orbital-excited $4f^8$ states, though, our atomic multiplet calculation does not provide the energy position of the $4f^7$ and $4f^9$ multiplets relative to $4f^8$. The relative energy depends critically on core-hole screening, which is altered by adding or removing a localized 4*f* electron. We determined the effect on the core-hole screening by optimizing the energy shift to match the experimental data (see SM for details).

There are issues with the 5*d*-4*f* electron transfer interpretation, that we discuss in the following: First, only about 0.1% of the 5*d* electrons/holes at 4000 K (peak temperature from the 2TM calculations) will exhibit energies compatible with respective tunnelling processes. To cause the observed effects, this channel must be very efficient. This at least is possible as the coupling between 5*d* and 4*f* states is rather strong (*32*). Second, we observe no indications of 4*f*-5*d* transitions in the RIXS data. As stated above, however, the RIXS cross section for specific 4*f* multiplets strongly varies with X-ray energy. We may have just missed these excitations with the particular photon energy of 147.2 eV. Finally, the contribution of the $4f^9$ seems to increase more than that of the $4f^7$ when going to longer delays. We speculate that this may be related to the different 5*d* density of states below and above $E_F$, affecting the 5*d*-4*f* overlap. We note here again that there is no common way to describe both the 4*f* and 5*d* electronic structure of RE metals in the same model and that a quantitative calculation of *d*-*f* scattering and electron transfer is currently beyond the state-of-the-art.

For Gd, which we used for a reference XAS measurement (see SM, Fig. S1), all the excitations discussed for Tb would occur at much higher energies; the lowest possible excitation out of the $^8S_{7/2}$ $4f$ multiplet requires an energy of 4.1 eV (*29*). The fact that we observe no spectral change in Gd XAS shows that all changes seen in Tb XAS are related to $4f$ electronic changes and not caused by other pump effects.

**Discussion**

Transient angular momentum change

While open questions remain for the excitations of higher $4f^8$ multiplets and the $5d$-$4f$ electron transfer, the XAS and RIXS data show the $4f^8$ $^7F_6 \rightarrow {}^7F_5$ multiplet excitation without any doubt as the dominant effect. This helps understanding angular momentum transfer in non-equilibrium magnetization dynamics of Tb and other RE ions. $4f$-state excitations need to be included in a full description of non-equilibrium processes, particularly because they could affect the coupling between the $4f$ system and the lattice, an important channel in magnetic dynamics of RE metals (*3,7*).

The effect of $4f$ orbital excitation goes beyond a mere rescaling of the $4f$-phonon coupling. We validate this effect, by investigating the influence of the $4f^8$ $^7F_5$ excited states on the MCA. At room temperature in the paramagnetic phase of Tb all $m_J$ levels ($J = -6 \ldots +6$) of the $^7F_6$ multiplet are occupied (*17*). In the ferromagnetic phase at low temperatures the $m_J$ level splitting increases and the energetically lowest configuration $m_J = 6$ ($m_\ell = 3$, $m_s = 3$) becomes preferentially occupied. To estimate the MCA change, we compare a Tb $m_\ell = 3$ ground state with an $m_\ell = 2$ state, which (together with spin-flip $m_s = 2$ states) forms the lowest $m_J = 5$ level of the $4f^8$ $^7F_5$ multiplet: $|m_J = 5\rangle = 1/\sqrt{2}\, |m_\ell = 2, m_s = 3\rangle + 1/\sqrt{2}\, |m_\ell = 3, m_s = 2\rangle$. From the $m_s = 2$ contributions, we expect a minor impact on the MCA, but a considerable one from $m_\ell = 2$. We therefore determined the MCA for the Tb $m_\ell = 3$ ground state and the $m_\ell = 2$ excited state from all-electron first-principles calculations (see Materials and Methods). For the ground state our *ab initio* calculations provide an MCA of 15 meV per atom with a preferential orientation of magnetic moments (easy axis) in the hcp basal plane along the crystallographic *a* axis, both of which are consistent with experiment (*33*). When we modify the minority Tb $4f$-electron occupation to the $m_\ell = 2$ state and perform a one-step calculation of the total energies, we find a rotation of the easy axis: magnetization along the *c* axis perpendicular to the basal plane becomes favourable. We verified this result by creating a self-consistent state with dominant $m_\ell = 2$ occupation by modifying the Coulomb repulsion between the $4f$ electrons (adjustment of Hubbard U), which, likewise flips the easy axis.

While we could prove the $^7F_6 \rightarrow {}^7F_5$ transition, the data indicate additional $4f$ excitations with minor weight and a time-dependent contribution of different excited states. In addition to the changes of the MCA, also contributions to ultrafast demagnetization could matter, e.g., by thermal occupation of the $4f$ states or via electron-transfer transitions. The latter could principally quench the MCA by orders of magnitude by creating Gd-like configurations (*7*).

Our work stands apart from studies on $4f$ orthoferrites, where the rare-earth ions (R) are placed in non-centrosymmetric positions of the $RFeO_3$ crystal lattice (*34-38*). As a consequence of symmetry breaking, resonantly driven electric dipole transitions dominate the observed dynamics. A resulting change of MCA provokes reorientation of the Fe spins in the crystal lattice; a transition which is also induced by thermal heating (*39*). We report on an ultrafast change of $4f$ orbital states, based on inelastic electron-electron scattering between itinerant $5d$ and localized $4f$ electrons.

Perspectives

In the Tb XAS experiment we altered the 4*f* state of about 20 % of all atoms (compare best fit, blue solid lines in Fig. 2E), which is quite sizeable. With a simple estimate we find a transfer of roughly 10 % of the absorbed laser energy into the 4*f* system (see SM). The excitation density might be further enhanced and specific transitions selectively driven by resonant pump-laser excitation. Shaping of the valence band structure by combining, e.g., 3*d* and 4*f* metals in alloys or multilayers could also affect the efficiency and selectivity of the 4*f* electronic excitations.

Our findings are particularly relevant in view of recent reports on all-optical switching (AOS) in Tb-based compounds (*40-42*). Compared to AOS-materials comprising magnetically soft Gd, the strong MCA in Tb promises stable domains on nanometer scale. The dynamics, however, appear to be more intricate in Tb-based systems; the accepted model describing AOS in Gd-based compounds fails. Our conclusion about the suppressed 4*f* excitations in Gd and their dominant contribution in Tb, gives another context to the discussion on AOS in Tb based compounds.

Our findings identify a fundamental excitation mechanism, based on 5*d*-4*f* inelastic electronic scattering. This so far disregarded process is highly relevant for optically induced magnetization dynamics in 4*f* metals and their compounds, as it provides basic insights to atomic spin and orbital dynamics, as well as to non-equilibrium 5*d*-4*f* coupling mechanisms. The presented electronic scattering channel might be exploited to optically control material parameters in magnetically ordered metals and further correlated materials. The ultrafast handle to transiently manipulate MCA may be functionalized for, e.g., writing bits in high density magnetic storage devices.

## Materials and Methods

X-ray absorption at the SCS station of EuXFEL

We performed the X-ray absorption experiment at the EuXFEL SCS Instrument (*43*), making use of the high energy resolution ($E/\Delta E \approx 3500$) in the energy range around the Tb and Gd 3*d* to 4*f* ($M_{5,4}$) resonances. We recorded high-resolution X-ray absorption spectra using the X-ray gas monitor (XGM) for measurement of the incident intensity $I_0$ and the transmission intensity monitor (TIM) - a cwd diamond scintillator - to determine the sample transmission $I_T$. The XAS signals (Fig. 2A) are calculated via $-\log(I_T/I_0)$. The time-resolved XAS measurements were performed with 350 meV energy resolution. The 65 fs time resolution was determined by cross-correlation of optical pump and X-ray probe pulses. For longer intervals of data recording the time resolution was 100 – 200 fs (data shown in Fig. 5). For exciting the samples, we used 800-nm pulses ($h\nu$ = 1.55 eV) from the SASE3 PP laser system. The incident pump fluence was (10 ± 2) mJ/cm². The laser-spot size was (0.28 x 0.2) mm$^2$ and thus larger than the X-ray-probe beam of (0.1 x 0.1) mm$^2$ ensuring a nearly homogeneous excitation profile in the X-ray spot.

Resonant inelastic X-ray scattering at the PG1 beamline of FLASH I

We conducted the RIXS experiments at the PG1 - TRIXS permanent end station (*44*) (RIXS spectrometer for Time-resolved Resonant X-Ray Scattering) located at FLASH. The PG1 beamline can reach energies from 36 to 250 eV with high intensity and an energy resolution of about 0.07%, ideal for investigating RIXS at the $N_{4,5}$ resonance of Tb around 147 eV. The experiment was performed at a grazing incident angle of about 25° and photons were collected under 105° with respect to the sample surface. The outgoing photons are dispersed according to their energy by a grating spectrometer and recorded on an X-ray CCD camera. In focus the RIXS grating spectrometer provides a resolution of about 100 meV/px at 147 eV, corresponding to about 140 meV total resolution in the spectrum. Off-focus the resolving power decreases according to the dispersion and due to the refocusing off-axial parabolic mirror. This results in a variation of the energy-resolution across the probed energy-loss region. For the RIXS data shown in Figs. 2B & 2F the spectrometer focus was optimized at 2.5 eV energy loss. For pumping we used near-infrared-laser pulses of 1030 nm wavelength ($h\nu$ = 1.2 eV) and achieved a time resolution of 300 fs (*45*). The incident pump fluence was 20 mJ/cm². The laser-spot size was (150 x 150) μm$^2$ and X-ray probe beam on the sample was (20 x 40) μm$^2$, ensuring a nearly homogeneous pump profile within the probed area. To record time traces (Fig. 3) we used a delay hopping routine randomly switching between delays to exclude unknown cross correlations.

Statistical Analysis

For the XAS and RIXS signals from the pumped and unpumped sample, we take the standard error for absorption and scattered signal respectively. For the differential change of the signals (Figs. 2E, 2F & 3) we deduce uncertainties from error propagation. The uncertainties for the differential change of absorption in the time intervals I and II of the energy-delay map (Fig. 5B & 5C) stem from the standard deviation in each energy-delay-bin.

Samples

In the XAS experiment we studied polycrystalline Tb and Gd transmission samples of 10 nm thickness. To prevent oxidation, the RE metal was sandwiched between yttrium layers. The samples were grown on an aluminum heat sink on a silicon nitride membrane, resulting in the following structure Y(2)/RE(10)/Y(25)/Al(300)/SiN(100), where RE = Tb and Gd. Numbers in brackets give the layer thickness in nanometers. The Tb and Gd samples were measured at room temperature in the paramagnetic phase.
For the RIXS study we grew a thicker Tb film (40 nm) to maximize the number of absorbed photons and therewith the count rate in the RIXS signal. The film was grown by molecular beam epitaxy (MBE) on a W(110) single crystal, prepared and characterized by X-ray magnetic circular dichroism at the PM3 beamline of BESSY II (Helmholtz-Zentrum Berlin) (*46*). To prevent the Tb film from oxidation the sample was capped with a sputtered 4 nm layer of Ta resulting in the following sample composition: Ta(4 nm)/Tb(40 nm)/W(110). The RIXS measurements were performed in the paramagnetic phase of Tb at room temperature.

Atomic multiplet calculations of $M_{4,5}$ edge XAS and $N_{4,5}$ edge RIXS

Because of the localized character of the 4*f* states, the shape of the absorption multiplets can be simulated even for laser-excited samples (*27*). The atomic multiplet calculations are performed using the Quanty simulation package (*47-49*). Details on the quantum chemical treatment of the atomic multiplet calculations of the X-ray spectra can be found in Section 4 – 6 of the Supplementary Text as well as in the book by de Groot and Kotani (*27,50*). We do not simulate 4*f*-5*d* interaction; we calculate atomistic XAS and RIXS spectra for different 4*f* configuration and occupation, which are superposed to describe the spectral changes.
For our approach in treating the X-ray absorption cross-section, it is assumed that for 4*f* RE compounds the 4*f*-4*f* as well as the 3*d*-4*f*-two-particle interactions are most important for the description of the $M_{4,5}$ (3*d*) X-ray absorption spectrum. Due to the large wave-function overlap the dipole term is dominated by $4f^{n}$ to $3d^{9}\,4f^{n+1}$ transitions. The calculations are based on spherical wave functions; a good approximation as spin-orbit interaction dominates crystal field effects in RE 4*f* metals. The interactions between the 3*d* core and 4*f* states in these atomic multiplet calculations are explicitly taken into account via the so-called Slater-Condon parameters. The Slater-Condon parameters used in this work were taken from Theo Thole's multiplet extension (*51*) to the Cowan code (*52*), which underlies the CTM4XAS interface maintained by de Groot *et al.* (*50*). The complete set of values can be found in Tables S1-S3. To correct for the Hartree-Fock over-estimation of electron-electron interaction the Slater reduction factors were set to $F_{ff} = 0.61$, $G_{df} = 0.70$ and $F_{df} = 0.80$ for the $M_{4,5}$ edge XAS calculations.

For the simulation of RIXS at the $N_{4,5}$ edge, the Kramers-Heisenberg formula was applied. The scattering geometry was set to 105° emission and 25° incident angles according to the experimental set-up. The reduction factors were set to $F_{ff}^2 = 0.80$, $F_{ff}^4 = 0.91$ and $F_{ff}^6 = 0.91$, $G_{df} = 0.60$ and $F_{df} = 0.60$ for the $N_{4,5}$ edge calculations.

DFT calculations

We performed DFT calculations, using the full-potential linear augmented plane wave (FP-LAPW) method in the local spin-density approximation (LSDA), as implemented in the programs ELK (*53*) and WIEN2k (*54*). Spin-orbit coupling is crucial in 4*f* systems, where it is stronger than the crystal field, and has been included in the calculations. The full Brillouin zone has been sampled by about 2000 *k*-points. Strong electron-electron correlations present in the Tb 4*f* states were included in terms of the Hubbard correction *U* and Hund's parameter *J* (*55*). The DFT+U double counting was treated in the fully localized limit. For $U = 9$ eV and $J = 0.5$ eV we obtain the self-consistent ground state Tb configuration with $m_\ell = 3$ and 15-meV MCA. The Tb configuration with $m_\ell = 2$ is self-consistently obtained for $U = 4$ eV and $J = 0.5$ eV. In the calculations where the $m_\ell = 2$ configuration was enforced without changing *U*, we employed the magnetic force theorem to obtain the MCA through a one-step total energy calculation. For the MCA calculations an increased accuracy of the muffin-tin potential and charge density expansion into spherical harmonics was used, with $\ell_{max} = 14$. MCA calculations for 4*f* based systems are often performed by mapping the full Hamiltonian to a crystal field (CF) Hamiltonian, where DFT calculations essentially provide the relevant CF parameters. The inaccuracies introduced by this mapping have been compared previously to LSDA based calculations (*56,57*), or DFT+U calculations similar to ours (*58*). These approaches involve different approximations, for example the CF mapping assumes that the crystal field is independent of the orientation of the 4*f* shell. On the other hand, the DFT+U calculations are highly sensitive to the choice of *U*.

**Acknowledgments**

The authors acknowledge European XFEL in Schenefeld, Germany, and DESY in Hamburg, Germany, for provision of X-ray free-electron laser beamtime at the Scientific Instrument SCS and the PG1 beamline of FLASH. We thank the instrument group and facility staff of SCS and PG1 for their assistance. Many thanks to BESSY II and staff for the opportunity to set up the MBE endstation at PM3 for sample preparation and characterization.

**Funding:**
This work was supported by:
Deutsche Forschungsgemeinschaft CRC/TRR 227 Ultrafast Spin Dynamics (Collaboration of projects A01, A03, A08 and Mercator Fellow)
Bundesministerium für Bildung und Forschung grant 05K19KE2
Swedish Research Council (VR)
Swedish Infrastructure for Computing SNIC grant 2018-05973
European Union’s Horizon2020 Research and Innovation Programme grant 863155 (s-Nebula)
K. and A. Wallenberg Foundation grant 2022.0079
Project Quantum materials for applications in sustainable technologies (QM4ST), funded as project No. CZ.02.01.01/00/22_008/0004572 by P JAK, call Excellent Research
Ministry of Education, Youth and Sports of the Czech Republic, project e-INFRA CZ (ID:90254)
Czech Science Foundation grant No. 23-04746S
Helmholtz Association grant VH-NG-1105

**Author Contributions:**
CS-L and MW declare equal contributions to this work.
Idea and proposal writing: NT-K, CSL
Sample preparation: TA, NT-K
Investigation at EuXFEL: NT-K, TA, WB, SJ, NP, RYE, PSM, MB, BE van K, MT, REC, LM, AY, GM, L Le G, NA, RG, MW, CS-L
Investigation at FLASH: NT-K, TA, WB, SD, GB, FP, RW, JOS, MS, GSC, MW, CS-L
Simulation of multiplet excitations: PSM, RW

DFT calculation: DL, KC, PMO
Two-temperature-model simulation: UA
Writing – original draft: NT-K, TA, MW, CS-L
Writing – review & editing: all authors

**Competing Interests:**
The authors declare to have no competing interests.

**Data and Materials Availability:**
All data needed to evaluate the conclusions in the paper are present in the paper and/or the Supplementary Materials.
Data recorded at the European XFEL is available at:
https://in.xfel.eu/metadata/doi/10.22003/XFEL.EU-DATA-002384-00
The Input file for the XAS atomic calculation performed with the Quanty simulation package is shown in the SI. The code for the DFT calculation is available from https://elk.sourceforge.io/ (Elk) and from http://susi.theochem.tuwien.ac.at (WIEN2k).
The data shown in this manuscript are available at:
https://zenodo.org/records/10575199

**Supplementary Materials**

Supplementary Text

Figs. S1 to S11

Tables S1 to S7

References (*59*-*69*)

**Figures**

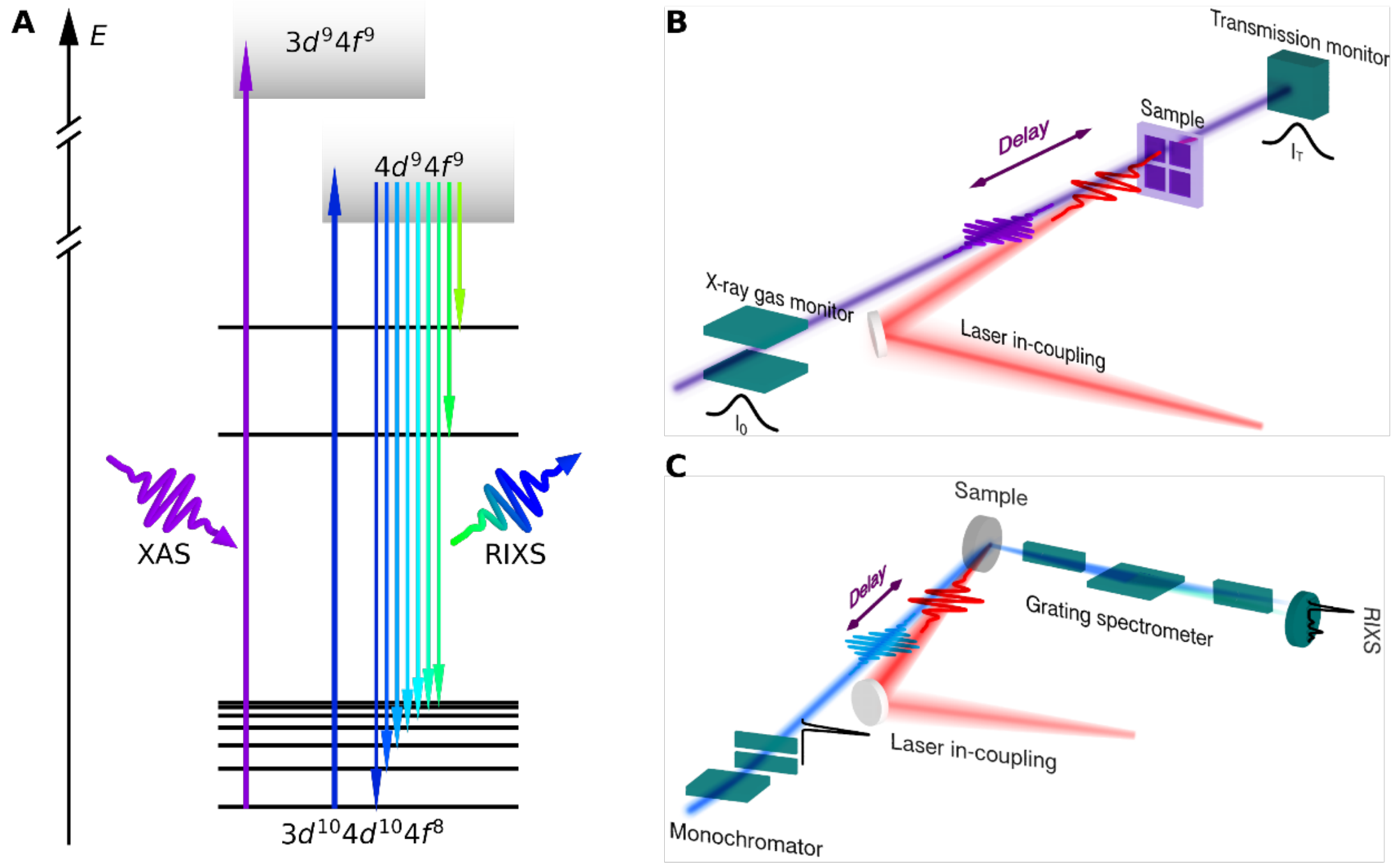

**Fig. 1. XAS Tb $M_5$ and RIXS Tb $N_{4,5}$ experiment.** (**A**) Scheme of the optical transitions in the X-ray absorption (XAS) and resonant inelastic X-ray scattering (RIXS) experiment. The black lines illustrate the energy levels in the $4f^8$ multiplet. (**B**) Sketch of the XAS experiment. Following the excitation with 1.55-eV photons, the transient absorption $log(I_T/I_0)$ is probed at the Tb $M_5$ edge in transmission geometry with 350 meV energy and 65 fs time resolution. The X-ray gas monitor is used to measure the incident intensity $I_0$. With the transmission intensity monitor we determine the sample transmission $I_T$. (**C**) Sketch of the RIXS experiment. After exciting the sample with 1.2-eV-laser pulses, the RIXS signal is probed with X-ray pulses at the Tb $N_{4,5}$ resonance and with a time resolution of 300 fs. With the TRIXS grating spectrometer we achieve an energy resolution of 140 meV.

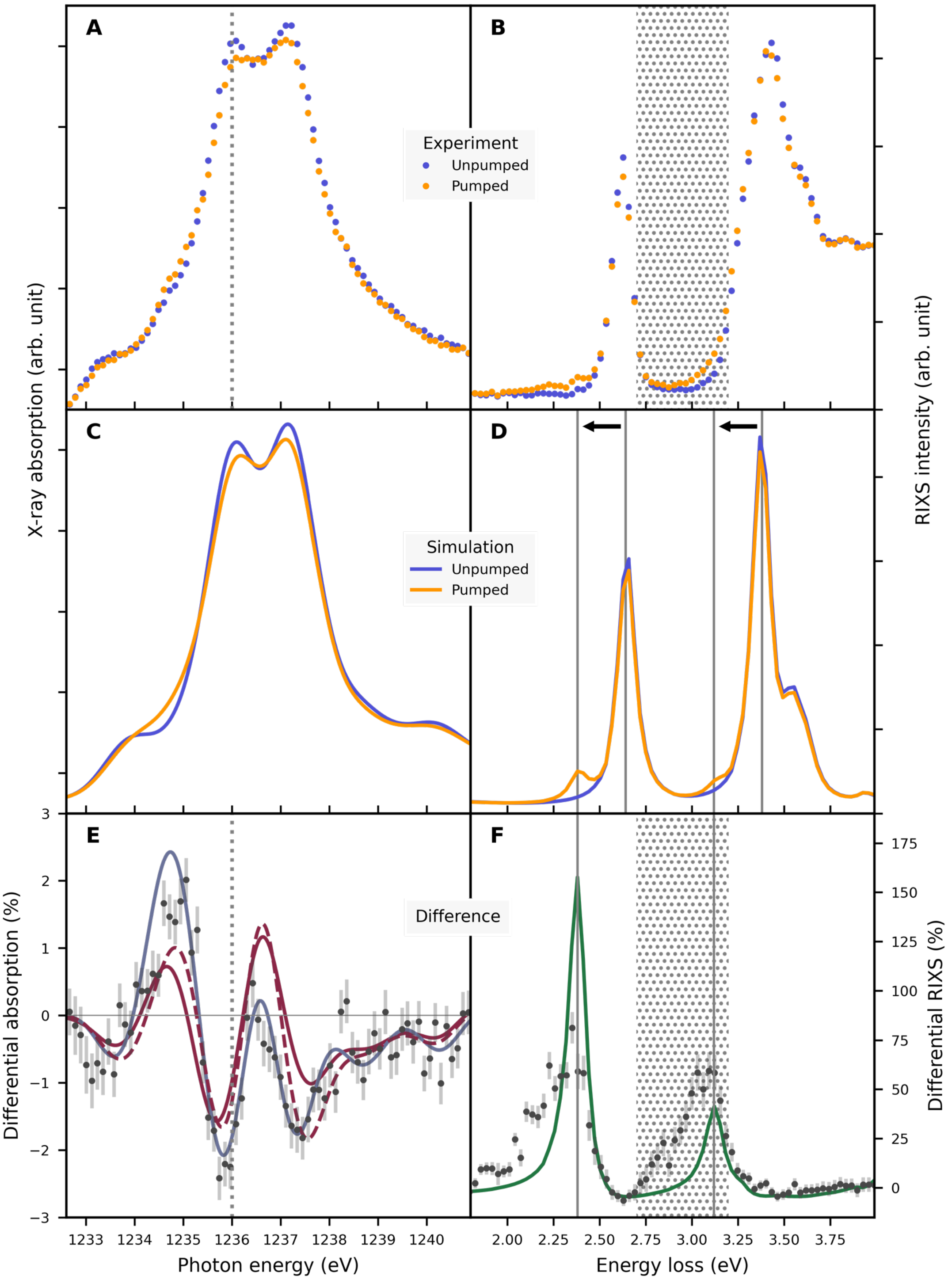

**Fig. 2. Pump-induced changes in the XAS Tb *M*5 and RIXS Tb *N*4,5 spectra.** (**A**) XAS spectrum of the Tb $M_5$ edge 150 fs after optical excitation (orange) in comparison to the spectrum for the unpumped sample (blue). The vertical line indicates the energy at which we recorded the pump-probe-delay trace (Fig. 3). (**B**) $N_{4,5}$ edge RIXS signal measured with 147.2-eV pulses for the unpumped Tb sample (blue) and 300 fs after pump-pulse excitation (orange). The hatched area marks the energy-loss window, over which we integrated the data to study the pump-probe-delay dependence (Fig. 3). (**C**) Simulation of the Tb $M_5$ spectra based on atomistic calculations. The blue line shows the pure Tb $4f^8$ $^7F_6$ ground state spectrum. The orange line shows the ground state spectrum, including admixtures of the excited states $4f^7$ $^8S_{7/2}$, $4f^8$ $^7F_5$ and $4f^9$ $^6H_{15/2}$. (**D**) Atomistic RIXS calculations for the pure $4f^8$

$^7F_6$ ground state (blue line) and with admixture of the $4f^8$ $^7F_5$ states (orange line). Horizontal arrows depict the shift of the main features by 0.26 eV ($^7F_6 \rightarrow$ $^7F_5$). (**E**) Differential X-ray absorption, i.e., the relative change of the XAS signal between pumped and unpumped sample (black dots). Solid lines are fits based on atomistic calculations of the $^7F_6$ ground state with admixtures of $4f^8$ $^7F_5$ (purple), and $4f^7$ $^8S_{7/2}$, $4f^8$ $^7F_5$, $4f^9$ $^6H_{15/2}$ (blue). The dashed purple line is a fit assuming thermal occupation of the $4f^8$ multiplets. The vertical line indicates the energy at which we recorded the pump-probe-delay dependence. (**F**) Differential RIXS, i.e., the relative change of the RIXS signal between pumped and unpumped sample (black dots). The green line depicts the relative change between the calculated spectra in (D). Pump-probe delay dependent data in Fig. 3 stem from integration over the energy-loss window marked by the hatched area.

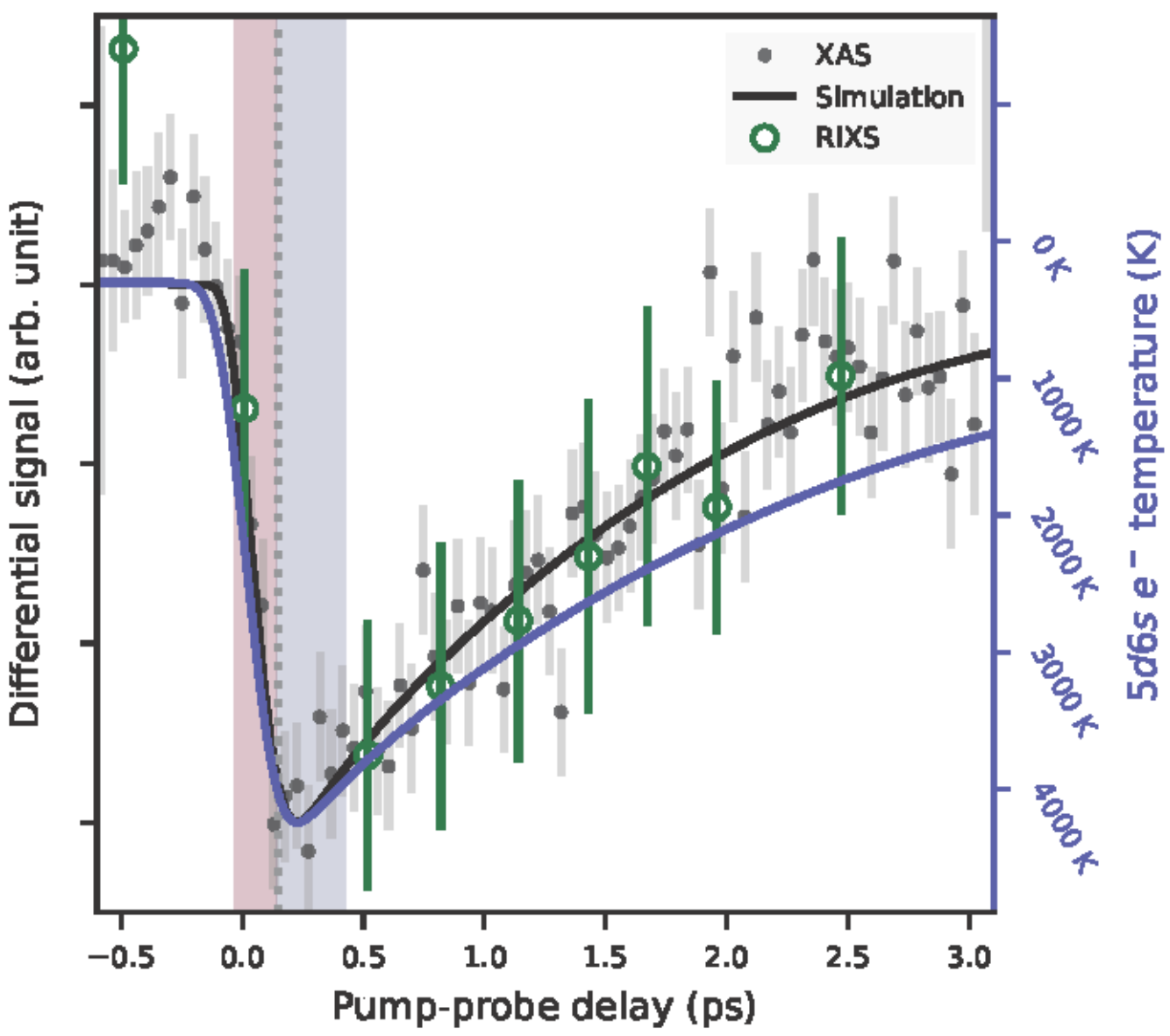

**Fig. 3. Pump-probe delay dependent changes in XAS Tb $M_5$ and RIXS Tb $N_{4,5}$ spectra.** Differential XAS (black dots) and RIXS signal (green markers) as a function of pump-probe delay. The XAS signal was measured at a photon energy of 1236 eV (vertical lines in Figs. 2A & 2E). The RIXS signal is integrated over the energy loss region 2.7 – 3.2 eV as marked in Figs. 2B & 2F. For comparison with XAS data the y-scale is inverted. The 5*d*6*s* electron temperature $T_{el}$ (blue solid line) has been calculated by the two-temperature model (see SM). The black solid line depicts the temperature evolution of those electrons, that can actually induce the 4*f* excitation observed in XAS. The colored bars indicate the time intervals over which we integrated the differential absorption shown in Figs. 5B & 5C.

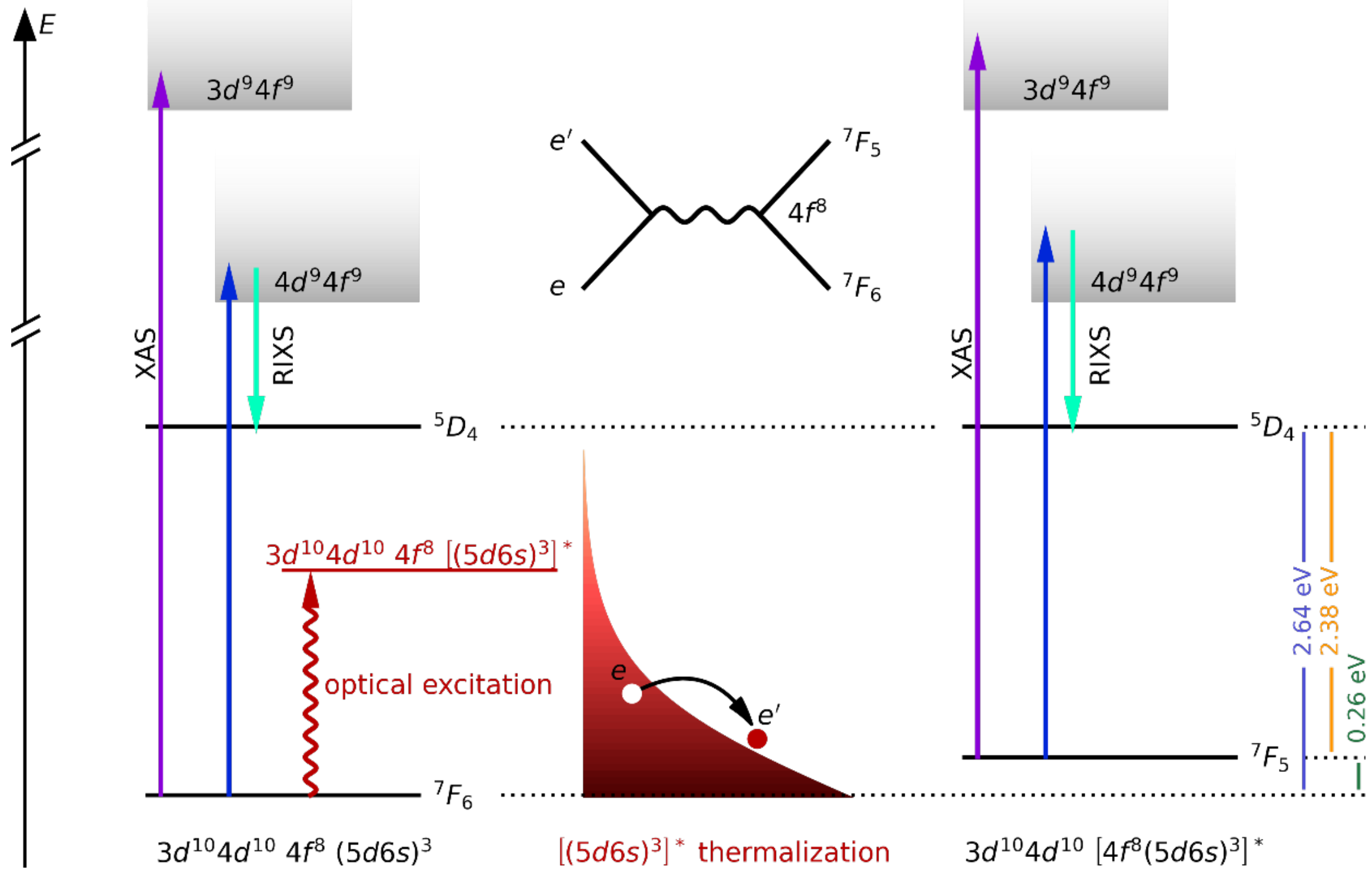

**Fig. 4. Total-energy scheme of the transitions involved in the XAS and RIXS experiment.** Left section: $3d$ ($4d$) electrons are excited at the $M_5$ ($N_4$) resonance in XAS (RIXS), provoking a transition from the initial state $3d^{10}4d^{10}4f^8(5d6s)^3$ to $3d^9 4f^9$ final ($4d^9 4f^9$ intermediate) states (purple [blue] arrows). In the initial state the $4f^8$ electrons occupy the $^7F_6$ ground state multiplet. The intermediate state in RIXS decays to a final state involving the $^5D_4$ multiplet (energy loss = 2.64 eV, compare RIXS spectra in Fig. 2B). Optical excitation by the pump laser elevates the full electron system from the ground state (red arrow) by exciting the $(5d6s)^3$ electrons. Middle section: The hot $[(5d6s)^3]^*$ electrons rapidly thermalize to a Fermi distribution (red area). Via inelastic electronic $5d$-$4f$ scattering (Feynman diagram) energy and angular momentum is transferred between the $(5d6s)^3$ and $4f^8$ electronic system. This leads to a $^7F_6 \rightarrow {}^7F_5$ transition (0.26 eV) with the $5d$ electron filling a hole of lower energy (– 0.26 eV). Right section: When the initial state for RIXS has changed to the higher lying $^7F_5$ multiplet, the RIXS feature associated with the decay to $^5D_4$ shifts by 0.26 eV to smaller energy losses. The XAS spectral shape changes compared to that from the $^7F_6$ state (Fig. 2C and Fig. S4 in the SM).

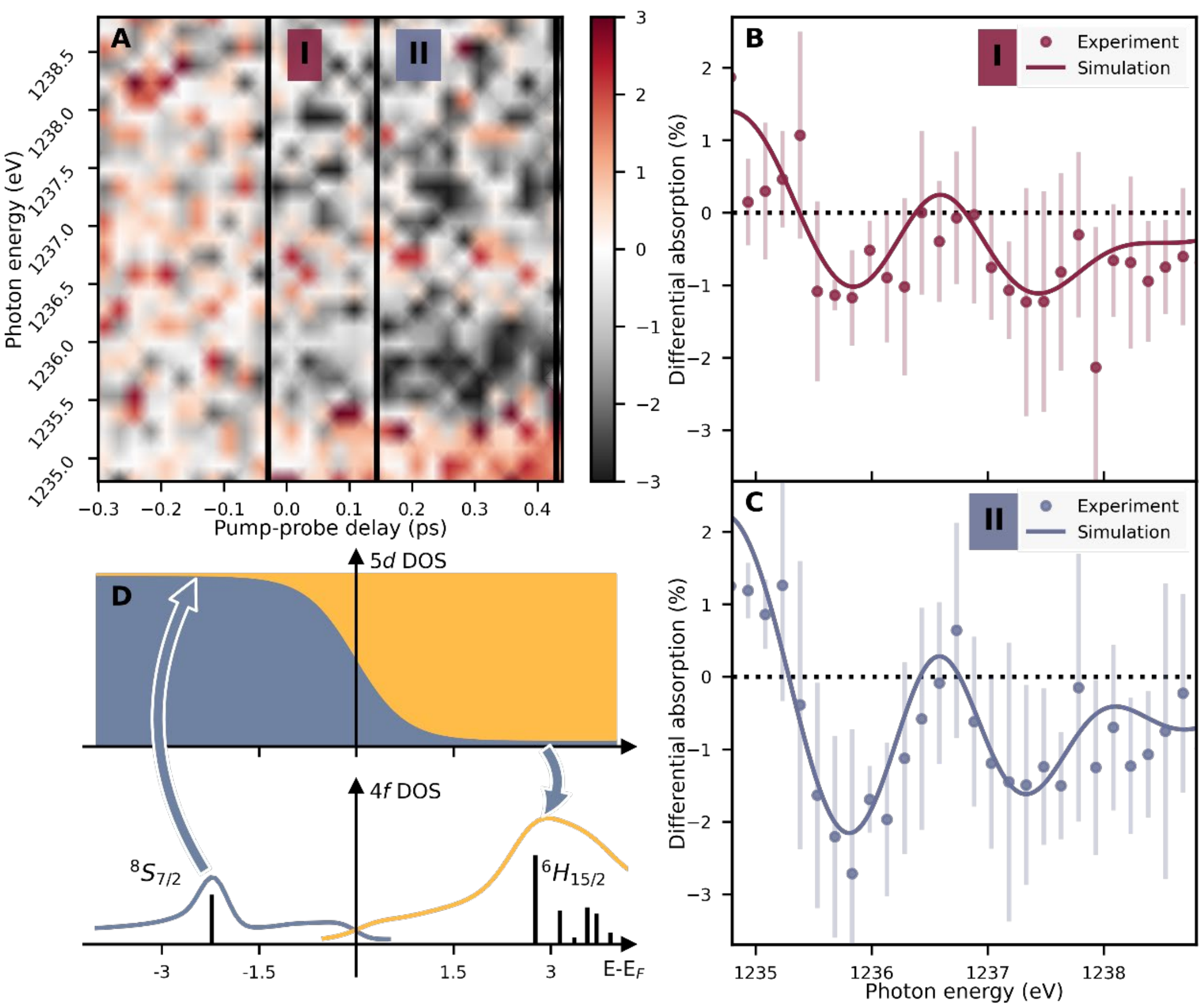

**Fig. 5. Pump-probe delay dependent variations in the Tb $M_5$ excitation multiplet.** (**A**) Map of differential X-ray absorption over photon energy (Tb $M_5$-edge) and pump-probe delay. We integrated the data in the time intervals I and II (black vertical lines, color-coded in Fig. 3) and obtained the results shown in (B) and (C). (**B**) From the map in (A) deduced energy-dependent differential absorption (dots) integrated over interval I (–0.03 ps to 0.143 ps). The solid line is a simulation including $4f^7$ and $4f^9$ electron-transfer contributions ($4f^7$ $^8S_{7/2}$ (2 ± 0.2 %), $4f^8$ $^7F_5$ (11.3 ± 0.8 %) and $4f^9$ $^6H_{15/2}$ (1.2 ± 0.5 %)). (**C**) From the map in (A) deduced energy-dependent differential absorption (dots) integrated over interval II (0.143 ps to 0.43 ps). The solid line represents a simulation including increased $4f^7$ and $4f^9$ contributions compared to the simulation shown in (B) ($4f^7$ $^8S_{7/2}$ (2.8 ± 0.2 %), $4f^8$ $^7F_5$ (19.7 ± 0.8 %) and $4f^9$ $^6H_{15/2}$ (4.8 ± 0.4 %)). (**D**) Within 100 fs after optical excitation 5*d* valence electrons form a hot Fermi distribution (dark grey area). Besides $^7F_5$ contributions from inelastic 5*d*-4*f* scattering, 5*d*-4*f* electron transfer possibly leads to $4f^7$ and $4f^9$ final states. The lower panel shows the $^8S_{7/2}$ final state of photoemission at 2.3 eV below $E_F$ and the $^6H_{15/2}$ final state of inverse photoemission at 2.8 eV above $E_F$ (*29*).

# Supplementary Materials for

## Optical control of 4*f* orbital state in rare-earth metals

Nele Thielemann-Kühn[1]*, Tim Amrhein[1], Wibke Bronsch[1,2], Somnath Jana[3]†, Niko Pontius[3], Robin Y. Engel[4]‡, Piter S. Miedema[4], Dominik Legut[5,6], Karel Carva[6], Unai Atxitia[1,8], Benjamin E. van Kuiken[7], Martin Teichmann[7], Robert E. Carley[7], Laurent Mercadier[7], Alexander Yaroslavtsev[7,9], Giuseppe Mercurio[7], Loïc Le Guyader[7], Naman Agarwal[7§], Rafael Gort[7], Andres Scherz[7], Siarhei Dziarzhytski[4], Günter Brenner[4], Federico Pressacco[4], Ru-Pan Wang[4,11], Jan O. Schunck[4,11], Mangalika Sinha[4]¶, Martin Beye[4]‡, Gheorghe S. Chiuzbăian[10], Peter M. Oppeneer[9], Martin Weinelt[1]# and Christian Schüßler-Langeheine[3]#

Corresponding author: nele.thielemann-kuehn@fu-berlin.de

**The PDF file includes:**

Supplementary Text
Figs. S1 to S11
Tables S1 to S7
References (*59-69*)

## Supplementary Text

### 1. Reference measurement on Gd

In order to check that no other effects than 4*f* electronic excitations may have interfered in the Tb X-ray absorption spectroscopy (XAS) experiment (like an interaction of the sample with the X-ray pulses) we did a reference experiment on Gd, which has the same crystalline and valence band structure as Tb, but differs in the 4*f* occupation. The lowest *f-f* excitation changing *J* within the half-filled Gd $4f^7$ shell multiplet requires an energy of 4.1 eV. Similarly, an electron-transfer excitation of a 5*d* electron to reach the $4f^8$ state takes 4.3 eV and the excitation of an 4*f* electron into the 5*d* manifold would require 8.0 eV ($4f^6$ final state) (*29*). Hence all excitation energies are much higher in Gd than in Tb so that multiplet excitations and charge-transfer processes must become significantly weaker. Indeed, for Gd pumped with a fluence identical to Tb we see no change of the absorption spectrum. Within our experimental accuracy the two XAS spectra in Fig. S1 recorded before excitation with the pump pulse (unpumped) and averaged over a delay range of 0 – 200 fs are identical. We hence relate any effects seen in the Tb XAS signal to changes of the Tb 4*f* electronic structure.

### 2. Exponential fits to pump-probe delay dependent traces (XAS)

Exponential least square fits to the pump-probe delay dependent XAS data were performed with the python lmfit package.

As fit model the following function is used:

```
def doubleDecayConvRec(x, t0, tau1,  tau2, A, C, sigma, offs):
   term1 = expConvGauss(x-t0,tau1,A,sigma)
   term2 = expConvGauss(x-t0,tau2,-C,sigma)
   term3 = ABCHConvGauss(x-t0,A,C,sigma)

   return (term1 + term2 + term3 + offs)
```

with

```
def expConvGauss(x,tau,A,sigma):
   term1 = np.exp(-x/tau)*np.exp(sigma**2/(2*tau**2))*
            (special.erf((sigma**2-x*tau)/(np.sqrt(2)*sigma*tau))-1)

   return -A/2*(term1)
```

and

```
def ABCHConvGauss(x,A,C,sigma):
   term1 = -A/2*special.erf(x/(np.sqrt(2)*sigma))
   term2 = +C/2*special.erf(x/(np.sqrt(2)*sigma))

   return term1+term2+1-A/2+C/2 .
```

For the exponential fit to the pump-probe-delay dependent XAS data (Fig. 3) we find the curve shown in Fig. S2. Here the decay time constant is $\tau_1 = (0.07 \pm 0.03)$ ps and the recovery timescale is $\tau_2 = (2.8 \pm 1.1)$ ps.

## 3. Simulation of electron temperature using the two-temperature model

When a metallic thin film is subjected to a near-infrared-laser pulse, electrons are excited by the photon electric field. Initially, the absorbed energy is barely transferred to the lattice and consequently the electronic system heats up. The electron and phonon temperatures, $T_{el}$ and $T_{ph}$, are decoupled for up to several picoseconds until the electron-phonon interaction equilibrates the two heat baths. This is phenomenology well captured by the so-called two-temperature model (2TM), (*59,60*) which can be written as two coupled differential equations:

$$(1)\ C_{el} \frac{\partial T_{el}}{\partial t} = -g_{ep} \left(T_{el} - T_{ph}\right) - \nabla \cdot \kappa_{el} \nabla T_{el} + P_{laser}(t,z)$$

$$(2)\ C_{ph} \frac{\partial T_{ph}}{\partial t} = +g_{ep} \left(T_{el} - T_{ph}\right)$$

The parameters entering Eqs. (1) and (2) are material dependent. For Tb, we only consider the excitation of the 5*d*6*s* valence electrons and set the specific heat $C_{el} = \gamma_{el} T_{el}$, where $\gamma_{el} = 225$ J/(m³K²) (*61*). As for the phonon specific heat we use the Einstein model (*62*):

$$(3)\ C_{ph} = C_{ph}^{\infty} \left(\frac{T_E}{T_{ph}}\right)^2 \frac{e^{T_E/T_{ph}}}{(e^{T_E/T_{ph}} - 1)^2}$$

Where $T_E$ is the so-called Einstein temperature. The Einstein model is sufficiently accurate for temperatures higher than the Debye temperature, $T_D$. Both, Einstein and Debye models provide very similar values for $T_E = 0.75\ T_D$, where $T_D = 174$ K and $C_{ph}^{\infty} = 2.2 \cdot 10^6$ J/(m³K) (*60,63*). According to Ref. *31* the electron-phonon coupling is $g_{ep} = 2.5 \cdot 10^{17}$ J/(sm³K). For metallic layers, the heat dissipation at room temperature is usually dominated by electron transport. The thermal conductivity $\kappa_{el}$ temperature dependent, increasing for the transient, high electron temperature according to $\kappa_{el} = \kappa_0\, T_{el} / T_{ph}$ (*63*). The thermal conductivity is $\kappa_0 = 16$ J/(smK) (*60*) and the initial temperature is $T_0 = 300$ K.

Optical pump fluence XAS

The optical pump- and X-ray probe beams were nearly coaxial and normally incident on the sample. From measured beam power and spot size and the 10 kHz repetition rate we estimate an incident pump fluence of (10 ± 2) mJ/cm². The optical constants for crystalline Y ($n = 2.13$, $k = 2.67$) and Tb ($n = 2.47$, $k = 3.27$) are taken from Ref. (*64*). For an RE(10)/Y(∞) stack and normal light incidence we estimate an absorbed power of (37 ± 1) % in a (10 ± 2) nm RE layer which corresponds to an absorbed energy of (3.7 ± 1.6) kJ/cm³.

To correctly estimate the absorbed power, effects like reflection between the layers need to be taken into account as well. For an idealized sample Y(2)/ Tb(10)/ Y(20)/Al(300)/SiN(100) (numbers in brackets is thickness in nm) a total absorption of $A$ = 0.332 and a total reflection of $R$ = 0.66 can be simulated, using the program IMD by David Windt (*65*). This corresponds to an absorbed power of (15.6 ± 0.1) % in the Tb layer.

An experimental measurement of the sample reflectivity for 800 nm light showed only (49.8 ± 0.8) % reflection under normal incidence. Such a lower reflectivity corresponds to a higher absorption. Hence, the absorbed power in the Tb layer is corrected to (23.5 ± 0.4) %. This accounts to an absorbed energy of (2.4 ± 1.3) kJ/cm³.

Estimation of the absorbed power

In Eq. (1), $P_{\text{laser}}$ represents the absorbed power density by the electron system, coming from the pump pulse with

$$(4)\ P_{\text{laser}}\,(t,z) = \frac{AF_0}{d}\frac{\exp\left(-t^2/2\tau^2\right)}{\sqrt{2\pi}\tau}\exp\left(-\alpha_{\text{opt}}\,z\right)$$

The 2TM does not include electron thermalization. Therefore, the laser duration was set to $\tau = 85$ fs to describe the internal thermalization of the electron gas via electron-electron scattering (~200 fs FWHM). In Eq. (4), $F_0$ is the incidence fluence, $d$ layer thickness, and $A$ absorption coefficient. To account for ballistic electron excitation out of the probed volume, the laser attenuation is estimated to a significantly increased value of $\alpha_{\text{opt}}^{-1} = 40$ nm in Tb (*66*). We can define the maximum power absorbed, $P_0$, which happens at $t = 0$ and $z = 0$ (surface layer):

$$(5)\ P_0 = P_{\text{laser}}\,(t = 0, z = 0) = \frac{AF_0}{d\sqrt{2\pi}\tau}$$

The total absorbed energy density can be calculated by integrating $P_{\text{laser}}\,(t, z)$ over time. For example, at the surface $E_{\text{abs}} = \int P_{\text{laser}}\,(t, z)\,dt = AF_0/d$ .

To fit the transient XAS data (Fig. 3) we set $E_{\text{abs}}\,(z = 0) \sim 2.75$ kJ/cm. In the absence of any other dissipation channel the peak temperature can be estimated as

$$(6)\ T_{\text{el,peak}}^2 = T_0^2 + \frac{2AF_0}{d\gamma_{\text{el}}}\,.$$

For the case considered here $T_{\text{el,peak}} \sim 5200$ K. In the presence of electron-phonon coupling and thermal transport, the peak temperature reduces to $T_{\text{el,peak}} \sim 4300$ K, for the parameters considered here. The total absorbed energy per unit area reduces to the integration over the sample, from $z = 0$ to $z = d$ and over time.

$$(7)\ E_{\text{abs}} = \frac{AF_0}{d}\frac{\left(1-\exp(-\alpha_{\text{opt}}d)\right)}{\alpha_{\text{opt}}}\,.$$

Electron-phonon relaxation time

In a first step, one can estimate the electron-phonon relaxation time, $\tau_{\text{ep}} = C_{\text{el}}/g_{\text{ep}}$ from Eq. (1) by assuming that $C_{\text{el}}\,(T_{\text{el}}) \sim \gamma_{\text{el}}\,(T_0 + \Delta T_{\text{el}}/2)$. For $\Delta T_{\text{el}}\ /2 >> T_0$ we can write $\tau_{\text{ep}} = \gamma_{\text{el}}\,\Delta T_{\text{el}}/(2g_{\text{ep}})$. For fixed electron-phonon coupling $g_{\text{ep}}$ and Sommerfeld coefficient $\gamma_{\text{el}}$, the relaxation time is defined by the peak temperature. As the peak temperature increases, the electron-phonon relaxation time increases. Thus, for a lower laser power the recovery $\sim \gamma_{\text{el}}^{-1}$ is faster.

Comparison to XAS

In Fig. 3 we compare the temporal evolution of the change in 4*f* absorption at 1236 eV (right ordinate, black) with the electron temperature $T_{\text{el}}$ (left ordinate, blue) after laser excitation. We calculated $T_{\text{el}}$ with the two-temperature model described above. With an absorbed energy of about 2.75 kJ/cm$^3$ we reach a maximal $T_{\text{el}}$ of around 4300 K (Fig. S3). Comparing both signals, we see that the decay of $T_{\text{el}}$ is somewhat slower than the decay of the XAS difference signal. Besides limitation of our electron-temperature simulation, in particular the neglect of an energy transfer into the $4f$ system, this discrepancy can also be related to the fact that in XAS we see the electron temperature relative to the energy of the first excited multiplet state $^7F_5$. We assume that all $5d$ electrons, that can lose $\Delta E = E_1 - E_F = 280$ meV by occupation of a $5d$ hole can lead to a population of the lowest excited Tb $4f^8$ state at $E_1$ above $E_F$. We neglect a variation of the density of states and assume a constant DOS $D(E) = D$. Hence, the number of electrons is proportional to the convolution of the Fermi functions for $5d$ electrons $[f(E,T_{el})]$ and $5d$ holes 0.28 eV below the respective electrons $[1 - f(-\Delta E,T_{el})]$ from minus to plus infinity:

$$(8)\quad g(\Delta E,T_{\mathrm{el}}) = D\int_{-\infty}^{\infty} f(E,T_{\mathrm{el}})\cdot[1-f(E-\Delta E,T_{\mathrm{el}})]\,dE$$

where $f(E, T_{el})$ the Fermi distribution. Hence,

$$(9)\quad g(T_{\mathrm{el}}) = \left[D\,k_B\,T_{\mathrm{el}}\frac{\ln(\exp(E/k_B\,T_{\mathrm{el}})+1)-\ln(\exp(E/k_B\,T_{\mathrm{el}})+\exp(\Delta E/k_B\,T_{\mathrm{el}}))}{\exp(\Delta E/k_B\,T_{\mathrm{el}})-1}\right]_{-\infty}^{\infty}$$

where $\Delta E/k_B = 3133$ K. We find $g(T_{el}) = D\Delta E/(\exp(\Delta E/(k_B T_{el})) - 1)$. Since the scaling factor between the differential X-ray absorption signal $P(t)$ and $g(T)$ is unknown we normalized $g(T_{el})$ such that $\Delta g_{max} = P_{max}$. In particular, we obtain $P = 1 - 0.03\, g(T) / \Delta g_{max}(T)$. The solid black line in Fig. 3 shows already a much better match to the XAS data applying this simple scaling to $T_e$.

## 4. Atomic multiplet calculations

The Schrödinger equation of a free atom contains the kinetic energy of the electrons ($p^2/2m$), the electrostatic interaction of the electrons with the nucleus ($Ze^2/r$), the electron-electron repulsion ($e^2/r$) and the spin-orbit coupling of each electron (l.s) (*67*):

$$(10)\quad H_{ATOM} = \sum_N \frac{p_i^2}{2m} + \sum_N \frac{-Ze^2}{r_i} + \sum_{pairs}\frac{e^2}{r_{ij}} + \sum_N \zeta(r_i)\,l_i\cdot s_i$$

The kinetic energy and the electrostatic interaction of the electrons with the nucleus are the same for all electrons in a given atomic configuration. They define the average energy of the configuration ($H_{av}$). The electron-electron repulsion and the spin-orbit coupling define the relative energy of the different terms within a configuration, for example leading to the Hund's rule assignment of the electronic ground state configuration. The electron-electron repulsion is very large, but the spherical average of the electron-electron interaction can be separated from the non-spherical part. The spherical average can then be added to $H_{av}$. Overall this $H_{av}$ is neglected in the atomic multiplet simulations of X-ray absorption: the difference between $H_{av}$ in the initial and final state of the X-ray absorption process defines the energy shift needed in the calculation to fit to the experimental absorption spectrum. The modified electron-electron Hamiltonian plus the spin-orbit coupling part determine the energies of the different terms within the atomic configuration in both the initial and final state of the X-ray absorption process. The terms of a configuration are indicated by their total orbital moment $L$, spin moment $S$ and total moment $J$ in a $^{2S+1}L_J$ format. The general formulation of the matrix elements of two-electron wave functions can be written as:

$$(11)\quad \left\langle {}^{2S+1}L_J\left|\frac{e^2}{r_{12}}\right|{}^{2S+1}L_J\right\rangle = \sum_k f_k F^k + \sum_k g_k G^k$$

$F^i(f_i)$ and $G^i(g_i)$ are the Slater-Condon parameters for, respectively, the radial (angular) part of the direct Coulomb repulsion and the Coulomb exchange interaction. The $f_i$ and $g_i$ are non-zero only for certain integer values of $k$ (running from 0 or 1 to $i$), depending on the configuration. The direct Coulomb repulsion $f_0$ is always present and the maximum value $i$ equals two times the lowest value of $\ell$. The exchange interaction Slater-Condon parameter $g_i$ is present only for electrons in different shells. In our case, the $M_{4,5}/N_{4,5}$ edge will be the $3d/4d$ and $4f$ states. For $g_k$, $k$ is even if $\ell_1 + \ell_2$ is even and $k$ is odd if $\ell_1 + \ell_2$ is odd ($\ell_1 = 2$, $\ell_2 = 3$ for $3d^9 4f^{n+1}$). The maximum value $i$ equals $\ell_1 + \ell_2$. A $4f^n$ configuration contains $f_0, f_2, f_4$ and $f_6$ Slater-Condon parameters, because $\ell = 3$ for a $4f$ electron, thus the maximum value is 6. The final state in X-ray absorption with the $3d^9\ 4f^{n+1}$ configuration contains $f_0, f_2, f_4, f_6$ (for direct $4f$-$4f$ interaction) and $f_2, f_4$ (for $3d$-$4f$/$4d$-$4f$ interaction) and $g_1$, $g_3$ and $g_5$ Slater-Condon parameters. The value for $f_0$ for the $4f^n$ initial state was based on the reduced $f_2, f_4$ and $f_6$ parameters:

$$f_0(ff) = (4/195) \cdot f_2 + (2/143) \cdot f_4 + (100/5577) \cdot f_6$$

The values for $f_0$ for the $3d^9 4f^n/4d^9 4f^n$ X-ray absorption final state (considering the interactions of $4f$-$4f$ electrons and $3d$-$4f$/$4d$-$4f$ electrons) were based on the reduced $f_2, f_4$ and $f_6$ and $g_1$, $g_3$ and $g_5$ parameters:

$$f_0(ff) = (4/195) \cdot f_2 + (2/143) \cdot f_4 + (100/5577) \cdot f_6$$

$$f_0(df) = (3/70) \cdot g_1(df) + (2/105) \cdot g_3(df) + (5/231) \cdot g_5(df)$$

The Slater-Condon parameters used in this work were taken from Theo Thole's multiplet extension (*51*) to the Cowan code (*52*), which underlies the CTM4XAS interface maintained by de Groot et al. (*50*). The complete set of values is listed in Table S1-S3 including the $4f$ spin-orbit coupling $\zeta$.

Correction of Slater-Condon parameters

The Hartree-Fock approximation only provides a very rough estimation of electron-electron interaction. It is usually overestimating the expectation energies. Therefore, we need to apply a Slater reduction factor to correct interaction strength. The reduction factor is not only depending on the probing edges but also influenced by the target sample. Thus, we set $G_{df} = 0.70$, $F_{df} = 0.80$ and $F_{ff} = 0.61$ for the X-ray absorption calculations at the $M_5$ edge and $F_{ff}^2 = 0.80$, $F_{ff}^4 = 0.91$ and $F_{ff}^6 = 0.91$, $G_{df} = 0.60$ and $F_{df} = 0.60$ for the $N_5$-edge calculations. The $g_1(df)$, $g_3(df)$ and $g_5(df)$ are multiplied by $G_{df}$, f2(df), f4(df) are multiplied by $F_{df}$ and the $f_2(ff)$, $f_4(ff)$ and $f_6(ff)$ for both the initial $4f^n$ configurations and $3d^9 4f^{n+1}$ configurations are multiplied by $F_{ff}^{(i)}$. Both the X-ray absorption and resonant inelastic X-ray scattering calculations were conducted by Quanty routine (*47-49*).

## 5. Calculation of X-ray absorption spectra at the Tb $M_{4,5}$ edge

Calculations were done with the so-called lua-input-files and an example of such an input-file is shown in Table S7. Fig. S4 shows calculated X-ray absorption spectra at the Tb $M_5$ and $M_4$ edges. The spectroscopic notation $^{2S+1}L_J$ describes the electron configuration of the $4f$ shell before the XAS probe step (without $3d$ core hole). The blue spectrum at the bottom of Fig. S4 shows excitations from the $3d_{5/2}$ and $3d_{3/2}$ core level to the $4f^8$ ground state (GS), i.e., the initial $^7F_6$ configuration at around 1235 and 1255 eV, respectively. We fitted the measured GS spectrum with the calculated GS spectrum with respect to energetic position and included an experimental Gaussian broadening. The six subsequent spectra are X-ray transitions to excited $4f$ states $^7F_J$ where $J = 5, 4, 3, 2, 1, 0$, respectively. The $^5D_4$ multiplet corresponds to the first spin-flip excitation within the $4f^8$ configuration. The energies required to excite these multiplets by $4f$ inner-shell transitions are listed in Table S4. We aligned all spectra on the global energy scale applying the shift determined for the GS.

The two spectra at the top of Fig. S4 describe X-ray transitions into the two lowest charge-transfer states. The $4f^7\ {}^8S_{7/2}$ configuration corresponds to the transfer of a $4f$ electron into an empty $5d$ state. This requires excited holes at an energy of about 2.3 eV below the Fermi level $E_F$, which corresponds to the lowest ionization energy of the $4f$ state, i.e. the single ${}^8S_{7/2}$ multiplet component in X-ray photoelectron spectroscopy (*29*). The $4f^9\ {}^6H_{15/2}$ configuration is reached by transfer of a $5d$ valence electron into the $4f$ shell. For this, electrons have to transiently occupy valence-band states at an energy of 2.8 eV above $E_F$, which corresponds to the lowest $4f$ excitation in bremsstrahlung isochromat spectroscopy (*29*).

## 6. Calculation of resonant inelastic X-ray scattering at the Tb $N_{4,5}$ edge

The Tb $N_{4,5}$ edge RIXS spectra were calculated using the Kramers-Heisenberg formula. The scattering geometry was set to 105° emission and 25° incident angles. Instead of the $3d$ core shell, the Tb $N_{4,5}$ edge couple with the $4d$ states. The ground state configuration is identical to the $M$ edge as indicated above since the core shell is filled. But the excitation energies are slightly altered since somewhat different Slater reduction factors were applied. The energies required to excite the multiplets by $4f$ inner-shell transitions are listed in Table S5. Fig. S5 (top panel) shows the calculated RIXS map at the Tb $N_5$ edges. The RIXS spectra in the bottom panel of Fig. S5 were extracted by averaging over the spectral range separated by horizontal lines in the RIXS map, respectively.

## 7. Simulation of X-ray absorption spectra at the Tb $M_5$ edge

In the X-ray absorption experiment we measured the intensity of the incoming X-rays $I_0$ and of the signal $I_T$ transmitted through the sample. As X-ray absorption we evaluate the extinction XAS = $-\log(I_T/I_0)$. For simulation of the pump-induced effect on the Tb $M_5$ absorption spectrum, as a first step the XAS signal in equilibrium for the unpumped sample is described by atomic calculations (see Section 4 and 5). The calculations deliver the imaginary part of the scattering amplitude $f$, which corresponds to the transition probability from the $3d_{5/2}$ ground state to the unoccupied $4f$ state with different $4f$ electronic configuration and on an arbitrarily shifted energy scale $e$. In the case of the *unpumped* signal we fit the $Im[f(e)]$ for the $4f^8\ {}^7F_6$ ground state (GS) multiplet to the XAS spectrum. This aligns the calculated and experimental energy scales, described by an energy shift $a$.

$$E(a) = e + a.$$

By convoluting $Im[f_{GS}(E)]$ with a Gaussian function we account for the energy resolution in the experiment and the core-hole lifetime broadening, which yields the absorption coefficient

$$\mu_{GS}(E) = \frac{4\pi}{\lambda} Im_{GS}(E) \otimes Gauss(\Delta E).$$

The XAS signal is

$$XAS_{GS} = \mu_{GS}(E) \cdot d + C + edge\ jump$$

where $d$ denotes the sample thickness and $C$ an offset. The *edge jump* is approximated as

$$edge\ jump = I_{ej}\ H(E_B) \otimes Gauss(\Delta E_{lifetime})$$

with $H(E_B)$ being a Heaviside-function at the binding energy $E_B$, scaled by $I_{ej}$ and core-hole life time broadened by convolution with a Gaussian function.

Simulation of the X-ray absorption spectrum for the Tb $4f^8$ ground state

For the fit parameters $a$, $\Delta E$, $d$, $C$, $I_{\text{ej}}$, $E_{\text{B}}$, $E_{\text{lifetime}}$ shown in Table S6 we find the description of the *unpumped* spectrum shown in Fig. S6, which is most accurate in the energy region of the two prominent peaks. We therefore concentrate on this energy range for the analysis.

Simulation of pump-induced spectral changes

For the simulation of the pump-induced spectral changes, the spectrum for the pumped sample is described by the GS spectrum with admixtures of different excited electronic configurations. Therefore, we write the absorption coefficient for the *pumped* spectrum as

$$\mu_{exc} = \left(1 - \sum\nolimits_i c_i\right) \mu_{GS} + \sum_i c_i \mu_i$$

where $i$ indicates the excited $4f$ state, $\mu_i$ the respective absorption coefficient and $c_i$ the relative contribution to the total absorption coefficient. The XAS signal from the pumped sample is

$$XAS_{exc} = \mu_{exc}(E) \cdot d + C + edgejump$$

with the fit parameters in Table S6 kept constant. The pump effect $P$ is described as differential X-ray absorption, i.e., as the relative change of the XAS signal in percent.

$$P = 100 \cdot \frac{XAS_{GS} - XAS_{exc}}{XAS_{GS}}$$

For the simulation $P$ is fitted with $c_i$ as fit parameters. In the case of $4f^8$ excited states contributing to the signal, we assumed a constant parameter $a$ (see Table S6) for the energy position of the multiplet. For $4f^9$ and $4f^7$ electronic configuration a correction of the energy $e_s$ is included, as the multiplets will appear at an altered energy position due to screening of the additional electron/hole in the $4f$ shell. Since the atomic calculations adequately describe the two main features in the GS spectrum at 1236 eV and 1237.2 eV, the fit is performed in the middle region of the spectrum (1234.8 eV – 1238.8 eV) denoted by the vertical dotted lines in Fig. S6.

Beginning with the simulation of the differential absorption for the dataset recorded at 150 fs pump-probe delay, as a first attempt we considered contributions, only from the lowest excited state $4f^8$ $^7F_5$. We find reasonable agreement for (16 ±1) % admixture of the $^7F_5$ multiplet to the GS, however do not achieve a quantitative description of the experimental data (Fig. S7).

Higher $4f^8$ multiplet excitations

The fit to the XAS data can be improved by including higher $4f^8$ excited states [$^7F_J$ with $J$ = (5,4,3,2,1,0)]. Therefore, we consider two different scenarios:

i. Assuming thermal equilibrium in the $4f$ electron system, we involve higher $4f^8$ multiplets with their occupation following a Boltzmann distribution. We further account for the degeneracy $2J+1$ of the multiplets. The contribution of a multiplet at $\Delta E_i$ is therefore set to $c_i = b_i(2J_i+1)/\Sigma_i(b_i(2J_i+1))$ with $b_i = \exp(-\Delta E_i/(k_B T_{4f}))$. The best fit is achieved for $T_{4f} = 1930 \pm 80$ K (see Fig. S8).
ii. We expect the $5d6s$ electronic system to be thermalized 150 fs after pump-pulse excitation and consider the contribution of $4f^8$ multiplets, lying at different energies $\Delta E_i$ above the ground state (see Table S4), to be linked to the amount of $5d6s$ electrons, that can transfer the required energy. The portion of respective $5d6s$ electrons, is described as a convolution of the Fermi functions for electrons at energy $E$ [$f(E,T_{5d})$] and $5d6s$ holes at energy $\Delta E$ below the respective electrons [$1-f(E-\Delta E,T_{5d})$], yielding [$f_i = \Delta E_i/(\exp(\Delta E_i/(k_B T_{5d})) - 1)$]. For the fit, we fixed the relative contribution of the $4f$ multiplets to the weight of the respective number of $5d6s$ electrons at a certain temperature $T_{5d}$. With a contribution of the multiplets at $\Delta E_i$ of $c_i = f_i/\Sigma_i f_i$, the best fit is achieved for $T_{5d} = 1110 \pm 30$ K (see Fig. S9).

We note, that ansatz i) and ii) yield some improvement of the fits. The respective temperature derived from the fits is much larger than the expected phonon temperature at 150 fs pump-probe delay, but smaller than the 2TM-deduced $5d6s$-electron temperature. Regarding the RIXS data, the asymmetric peak shape of the differential RIXS features (Fig. 2F) could come from higher lying excitations. However, the discontinuous variation of energy resolution, renders a quantitative analysis difficult. Additional RIXS data from the anti-Stokes side of the spectrum seem to show no indications for other contribution than the $^7F_5$ states (see Fig. S10). At present, the existence of higher $4f^8$ excitations cannot be unequivocally proven.

Electron transfer excitations

We observe an increase of the XAS pump effect at 1235.8 eV over time (Fig. 5). A different contribution of higher lying $4f^8$ multiplets to the spectral change do not explain this observation (here the pump effect at the second feature at 1237.5 eV is pronounced, see Fig. S11) we consider the energetically lowest $4f^7$ ($^8S_{7/2}$) and $4f^9$ ($^6H_{15/2}$) multiplets in the simulation. This improves the fit considerably.  The spectral shape itself can readily be simulated and experimentally observed in Gd and Dy, respectively. However, the atomic multiplet calculation depends critically on core-hole screening, which is altered by adding or removing a localized $4f$ electron. We dealt with that by estimating the relative energy positions from the energy separation between the multiplet terms after alignment of the calculated spectra at the first absorption peak assuming metallic screening. For metallic screening all $3d$ to $4f$ excitonic transitions should be equally well screened and the differential shifts correspond to the $f$-$f$ excitation energies of – 0.28 and +0.13 eV (Table S4). We used these shifts as starting values for our fit. For the fits in Fig. 2E and Figs. 5B & 5C of the manuscript, we arrived at slightly larger shifts of – 0.44 and +0.53 eV for the $4f^7$ and $4f^9$ states, respectively (total energy shift $e_s$ of the spectra, $4f^7$: ($-1.68 \pm 0.08$) eV, $4f^9$: (0.74 $\pm$0.04) eV).  The shifts deduced from fitting the data may be justified by stronger screening of the 3$d$ core hole by the localized 4$f$ as compared to the delocalized 5$d$ electrons. In that case the binding energy of the core exciton is higher in the $4f^7$ and lower in the $4f^9$ state.

Within these settings we find (72 ± 2) % of the $^7F_6$ GS, (20 ±1) % of the $^7F_5$, (4.5 ± 0.7) % $^6H_{15/2}$ for $4f^9$ and (3.3 ± 0.2) % $^8S_{7/2}$ for $4f^7$, contributing to the *pumped* absorption spectra at 150 fs pump-probe delay.

For the simulation of the pump effect obtained from cuts through the energy-delay map in Fig. 5A of the main text, we adopted the parameters from Table S6 and the energy correction values $e_S$. Considering $4f^8$ as well as $4f^9$ and $4f^7$ excitations we find the best fits shown in Figs. 5B & 5C. Here we use (2 ± 0.2) % $4f^7$ $^8S_{7/2}$, (11.3 ± 0.8) % $4f^8$ $^7F_5$ and (1.2 ± 0.5) % $4f^9$ $^6H_{15/2}$ multiplets in interval (I) – 0.03 to 0.14 ps and (2.8 ± 0.2) % $4f^7$ $^8S_{7/2}$ , (19.7 ± 0.8) % $4f^8$ $^7F_5$ and (4.8 ± 0.4) % $4f^9$ $^6H_{15/2}$ in interval (II) 0.14 to 0.43 ps.

**8. Evaluation of RIXS data**

After binning the RIXS data over delay, the averaged spectra for each delay-bin were background-subtracted by the pump-pulse background recorded for each run and the background of diffuse scattered light. Secondly the data were normalized to the probe-pulse intensity and to the integrated intensity of each spectrum in the range of + 2 eV to – 10 eV. This normalizes to the overall emitted light, making the spectra at different delays comparable.

**9. Time resolution of the XAS experiment**

The duration of the optical laser pulse was 30 fs and of the FEL pulse 25 fs. The measured arrival time jitter for small time spans of 10 min was on the order of 50 fs. Combining these numbers gives a temporal resolution of about 65 fs. This is the time resolution of the measurements presented in Fig. 2 of the main text. There were also slower drifts on the order of 100 – 200 fs on longer timescales. We cannot exclude that these shifts contribute to the recorded energy vs. delay map and differential absorption spectra in Fig. 5 of the main text.

**10. Samples**

For the XAS experiment we studied polycrystalline transmission samples of 10 nm thickness sandwiched between Y-layers. Samples were grown by molecular beam epitaxy in the combined MBE/SD (sputter deposition) chamber of the PM3 beamline at BESSY II. The base pressure in the chambers is 1 x $10^{-10}$ and 1 x $10^{-9}$ mbar for MBE and SD, respectively. As substrate we used an Al heat sink on a silicon nitride membrane. The pressure during evaporation was 7 x $10^{-9}$ mbar at maximum. The stack was Y(2)/RE(10)/Y(25)/Al(300)/SiN(100), where the number indicates the nominal layer thickness in nanometer and RE = Gd or Tb, respectively. Due to different positions of the evaporators with respect to the quartz microbalance, we estimate thickness variations on the order of 30 %.

The Tb sample used for the RIXS experiment was epitaxially grown in the MBE/SD preparation chamber, as well. Here we grew 40 nm Tb on a W(110) single crystal via MBE. For oxidation protection we capped the sample with 4 nm Ta using SD. The pressure during Tb growth was 2.7 x $10^{-9}$ mbar and Ta was sputtered at an Ar pressure of 2 x $10^{-3}$ mbar.

Both, MBE growth of RE metal on W and Y, are established methods. As for the different film thicknesses, it has been shown that Tb films grow in bulk like structure after a few deposited layers (*68*). We therefore expect no influence on the electronic structure due to the different film thicknesses.

## 11. Transfer of energy density

As for the XAS experiment we assume that maximal 20 % of all Tb atoms are in the first excited multiplet state, we calculate for a density of 3.12 x $10^{22}$ atoms/$cm^3$ a stored energy of 0.27 x $10^3$ J/$cm^3$ (as used for the simulation of the XAS data with the 2TM-model) which corresponds to about 10 % of the maximum absorbed energy density.

## 12. Two-photon absorption vs thermal occupation

Two-photon absorption can contribute to excitation of multiplets with excitation energy below $2\hbar\omega \sim 3.1$ eV, but only while the pump-pulse excites the sample, i.e., around zero delay. Independent of the initial distribution, electrons will still thermalize to a hot Fermi distribution within 200 fs. The temperature of this distribution decays as described in Fig. 3 of the main text and Fig. S3. From this description we expect the highest contribution of $4f^7$ and $4f^9$ multiplet excitations at around 150 – 400 fs where $T_{el} \sim 4000$ K. The probability to find an electron at $E - E_F = 2.8$ eV is on the order of $10^{-4}$. We note that it is not straightforward to estimate the contribution of two-photon absorption. While single photon absorption scales with the laser-pulse intensity, two photon excitations scale with the square of the intensity. Space charge effects due to the IR pump pulse ($h\nu = 1.6$ eV) in time-resolved photoemission of Gd set in at an absorbed fluence of 0.5 mJ / $cm^2$ (~ $10^{15}$ photons / $cm^2$) (*69*). In direct photoemission with VUV pulses, we detect in the same setup about 1 electron per $10^8$ photons / $cm^2$ and space charge sets in at only slightly higher photon flux. Thus, we argue that the ratio between two-photon photoemission and direct photoemission must be on the order of $10^{-6}$ for the fluence range we also used in the present experiment, i.e., an overall contribution of two-photon absorption in the order of 1%.

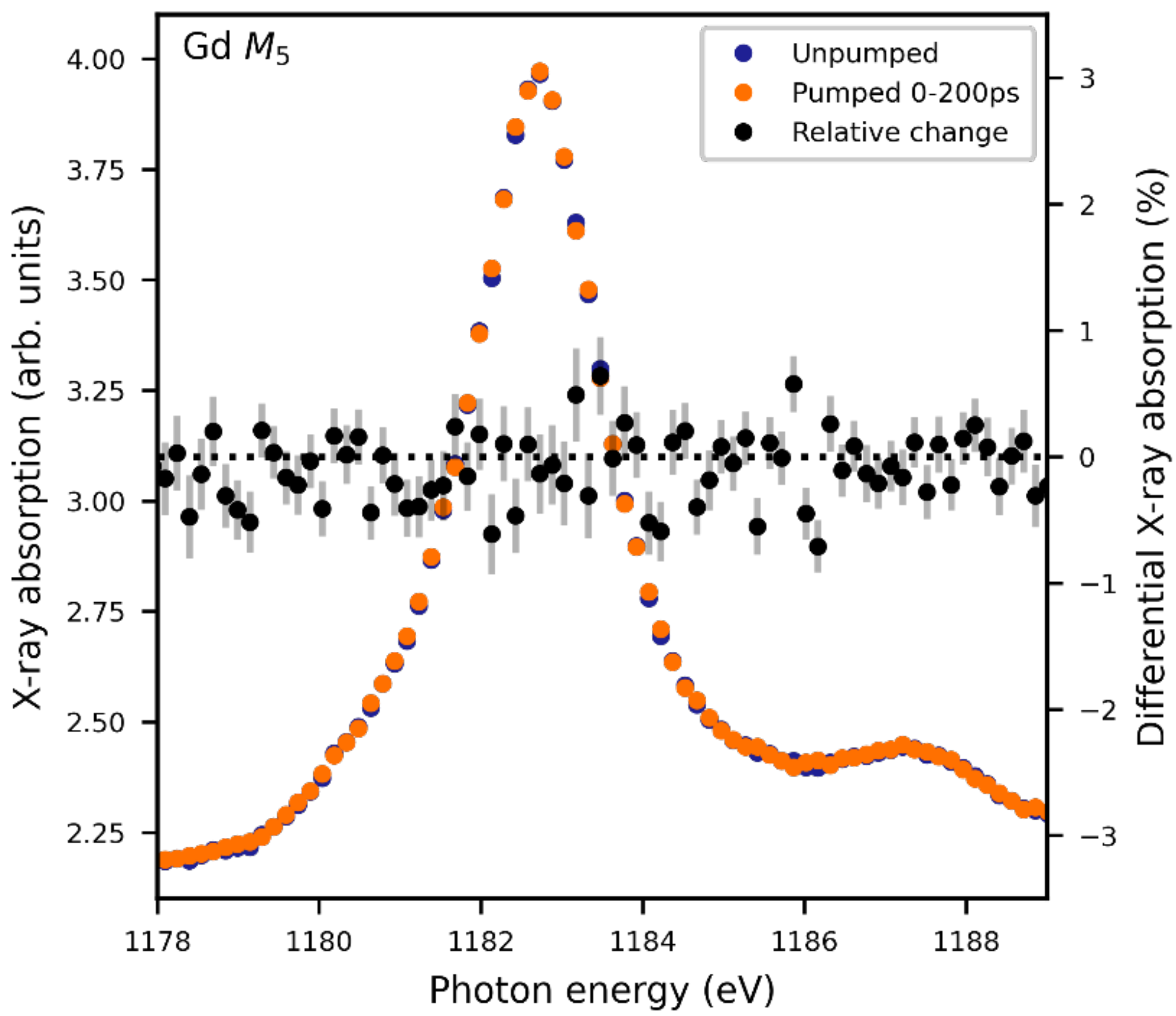

**Fig. S1.**
XAS spectrum of the Gd $M_5$-edge averaged over an interval of 0 – 200 fs after excitation with an 800-nm pump pulse (orange dots) in comparison to the spectrum for the unpumped sample (blue dots), and their relative difference on an enlarged vertical scale (black markers).

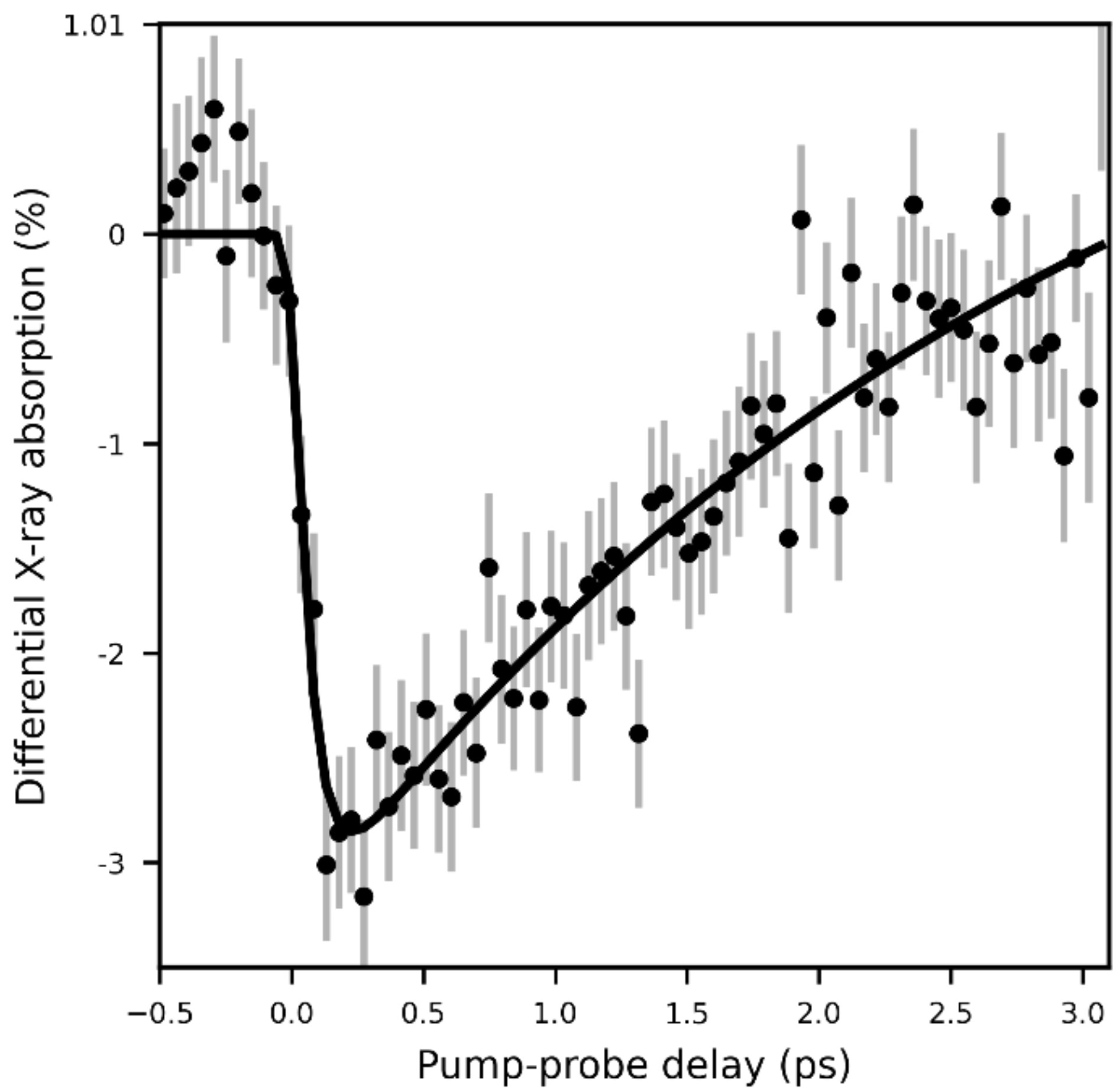

**Fig. S2.**
Differential XAS signal (black data points) as a function of pump-probe delay measured at a photon energy of 1236 eV (Tb $M_5$ resonance). The black solid line is an exponential least square fit to the data (see Section 2). The error bars origin from error propagation of the standard deviation of the measured signals $I_T$ and $I_0$.

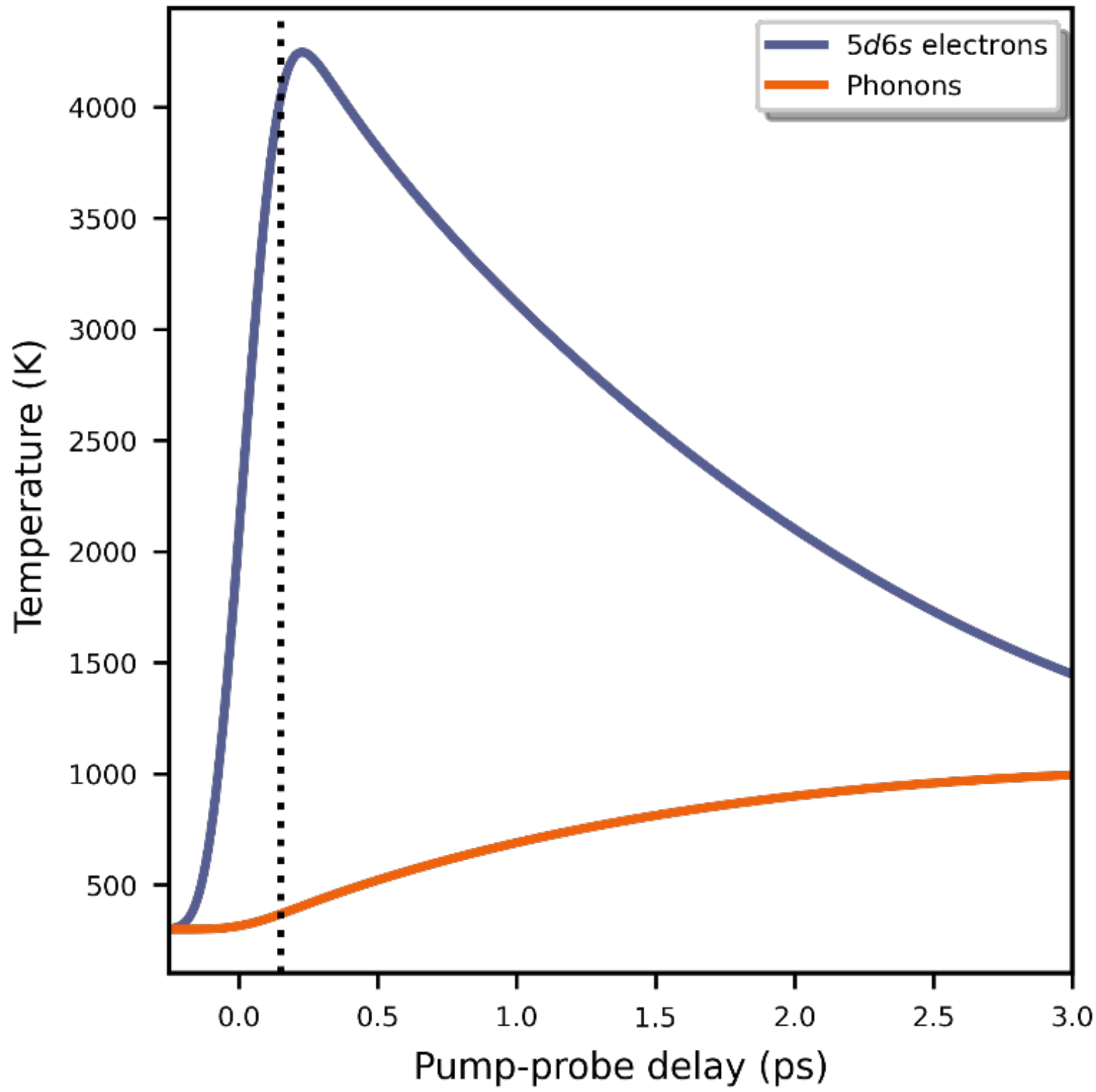

**Fig. S3.**
Evolution of 5*d*6*s* electron (blue) and lattice (orange) temperatures after pump-pulse excitation calculated according to the two-temperature model. The dashed vertical line indicates $T_{el}$ ~ 4000 K reached at a pump-probe delay of 150 fs.

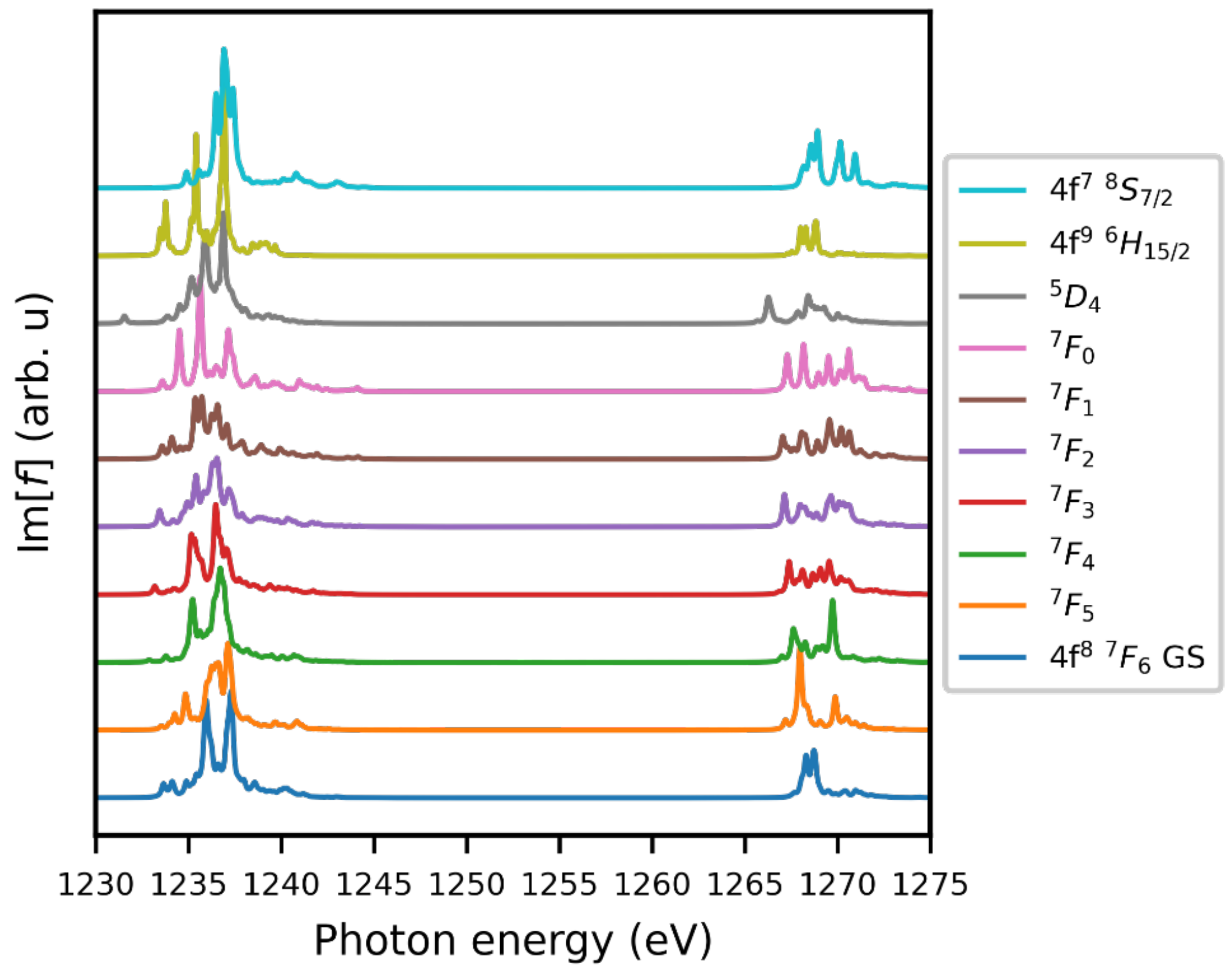

**Fig. S4.**

Calculation of multiplet transitions. The spectroscopic symbols denote the 4*f* configuration $^{2S+1}L_J$, where *S*, *L* and *J* are the spin, orbital and total angular momentum, respectively. Starting at the bottom, we show the Tb ground state $4f^8$ $^7F_6$, followed by higher $4f^8$ excitations ($^7F_J$ with $J = 4,3,2,1,0$, and $^5D_4$). Corresponding excitation energies are listed in Table S4. $4f^7$ and $4f^9$ are multiplet excitations after 5*d*-4*f* electron transfer.

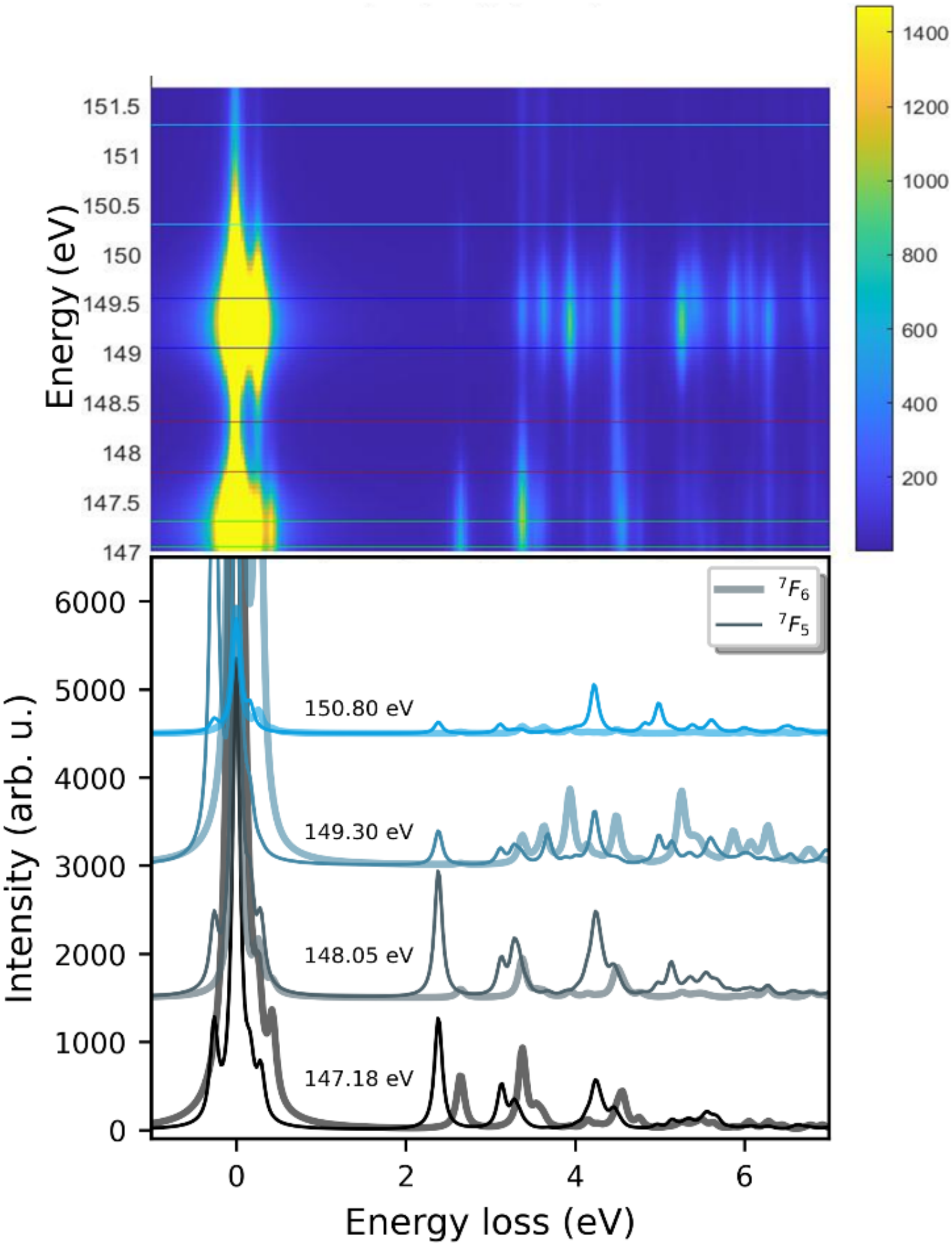

**Fig. S5.**

Calculation of the Tb $4f^8$ $^7F_6$ RIXS map at the $N_{4,5}$ resonance (top) and horizontal cuts through this map (RIXS spectra, bottom) for the $^7F_6$ (thick solid lines) and $^7F_5$ (thin solid lines) multiplets.

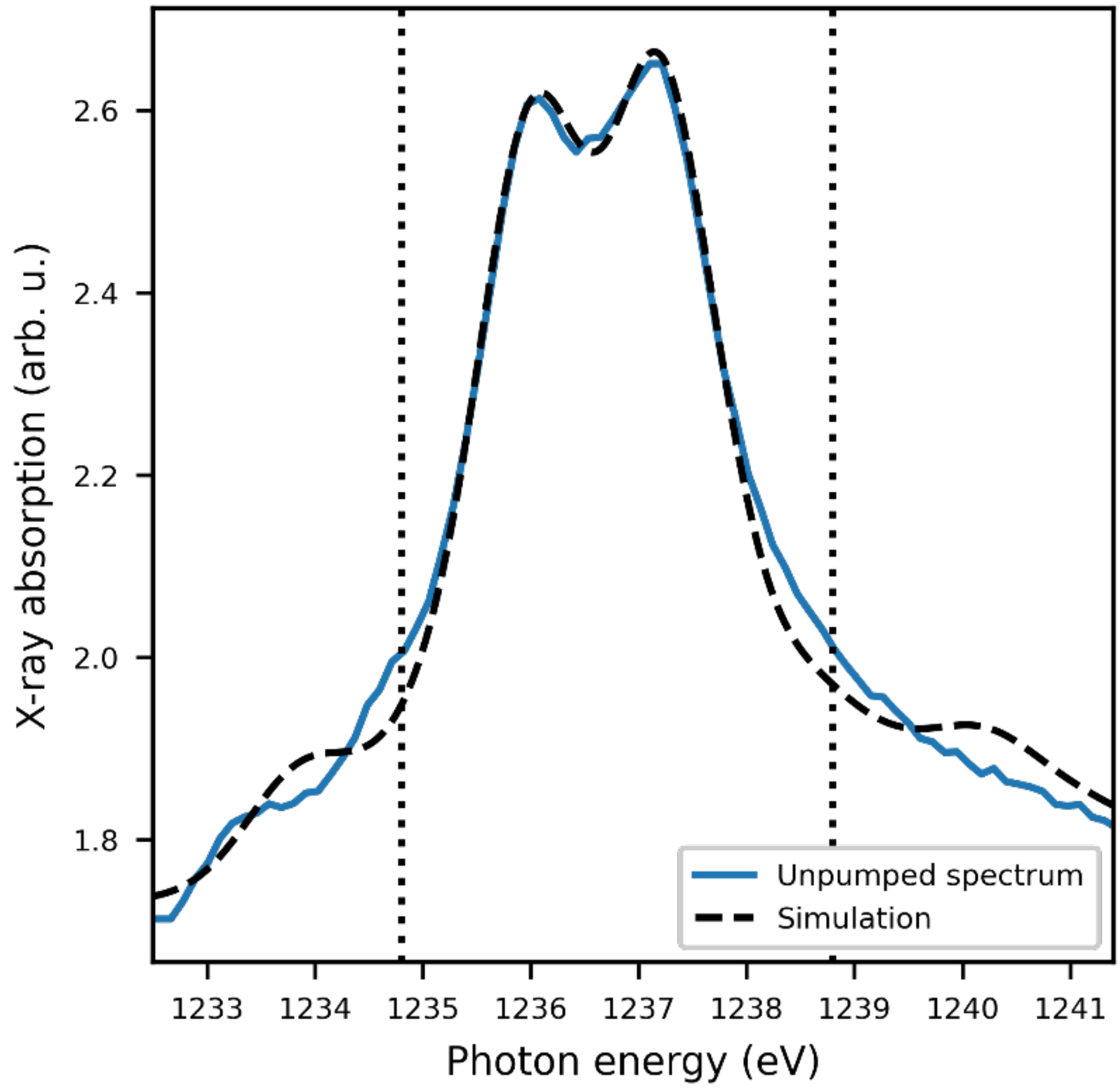

**Fig. S6.**
Simulation of the *unpumped* Tb $M_5$ XAS spectrum with the parameters shown in Table S6. The solid line (blue) marks the experimental XAS data and the dashed line (black) shows the simulation. We restrict the fit to the two main peaks in the energy range of 1234.8 to 1238.8 eV (dashed vertical lines).

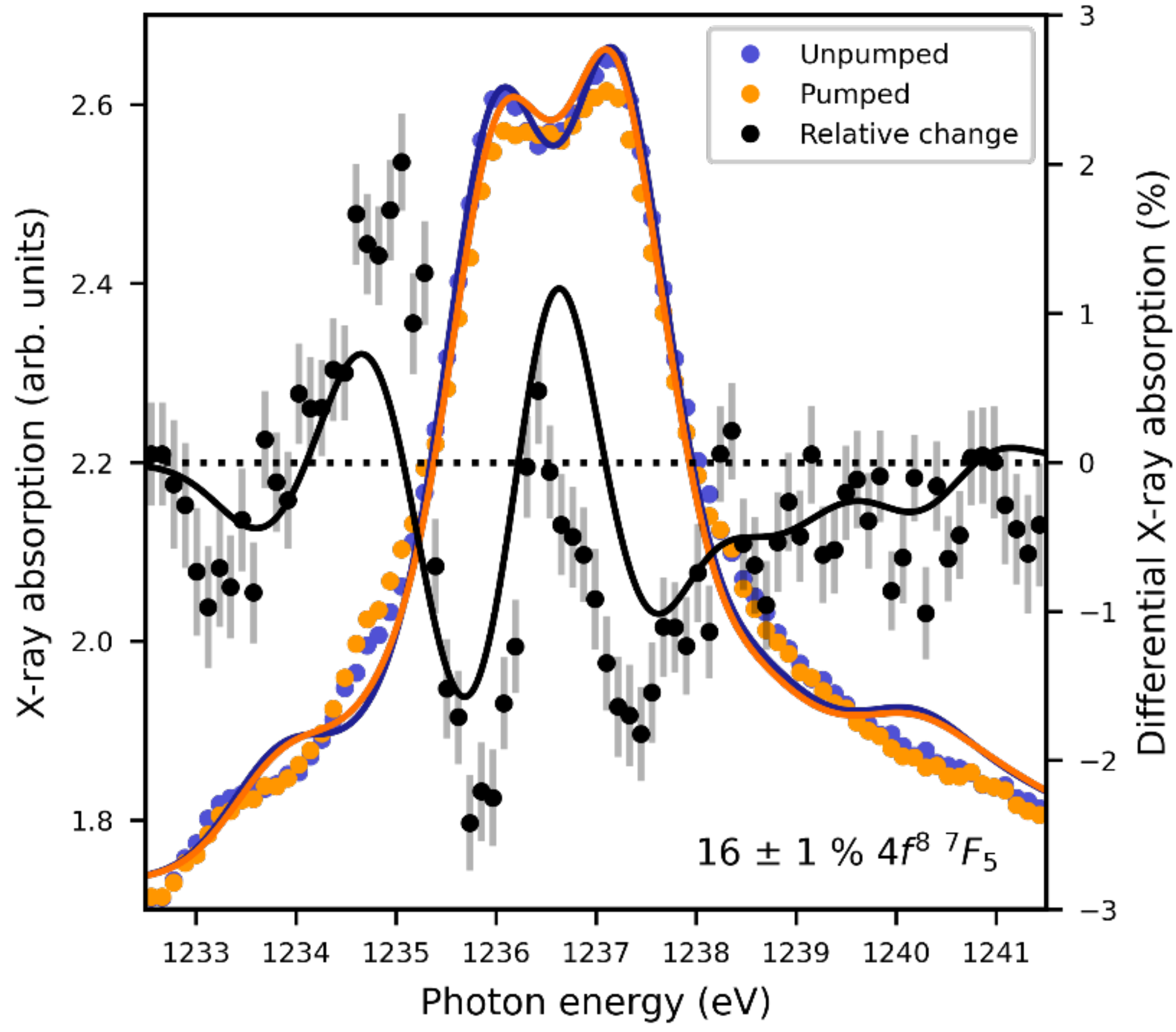

**Fig. S7.**
Tb $M_5$ resonance at 150 fs pump-probe delay for pumped (orange dots) and unpumped sample (blue dots), as well as the differential X-ray absorption plotted to the right abscissa (black dots). The error bars are deduced from the error propagated standard deviation of the measured transmission $I_T$ and $I_0$. The lines show the simulation of the data, based on a fit of the difference signal, assuming admixtures of the lowest $^7F_5$ multiplet to the $^7F_6$ ground state. We find best agreement for (16 ± 1) % $^7F_5$ and (84 ± 1) % $^7F_6$ contribution.

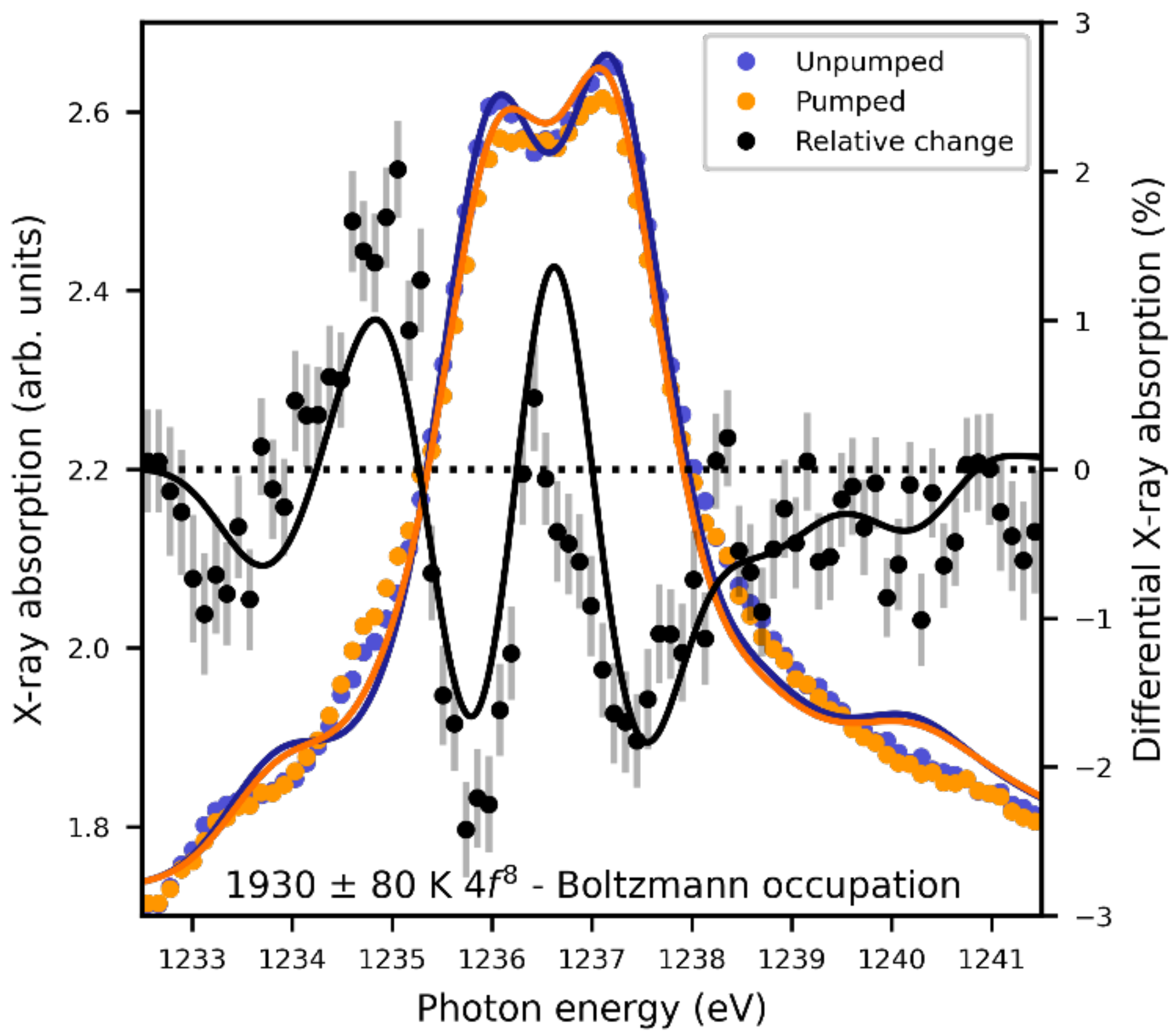

**Fig. S8.**
Tb $M_5$ resonance at 150 fs pump-probe delay, as well as the differential X-ray absorption (dots). The lines show the simulation of the data, based on a fit of the difference signal, assuming a $4f^8$ multiplet occupation following a Boltzmann distribution. We find best agreement for a temperature of (1930 ± 80) K.

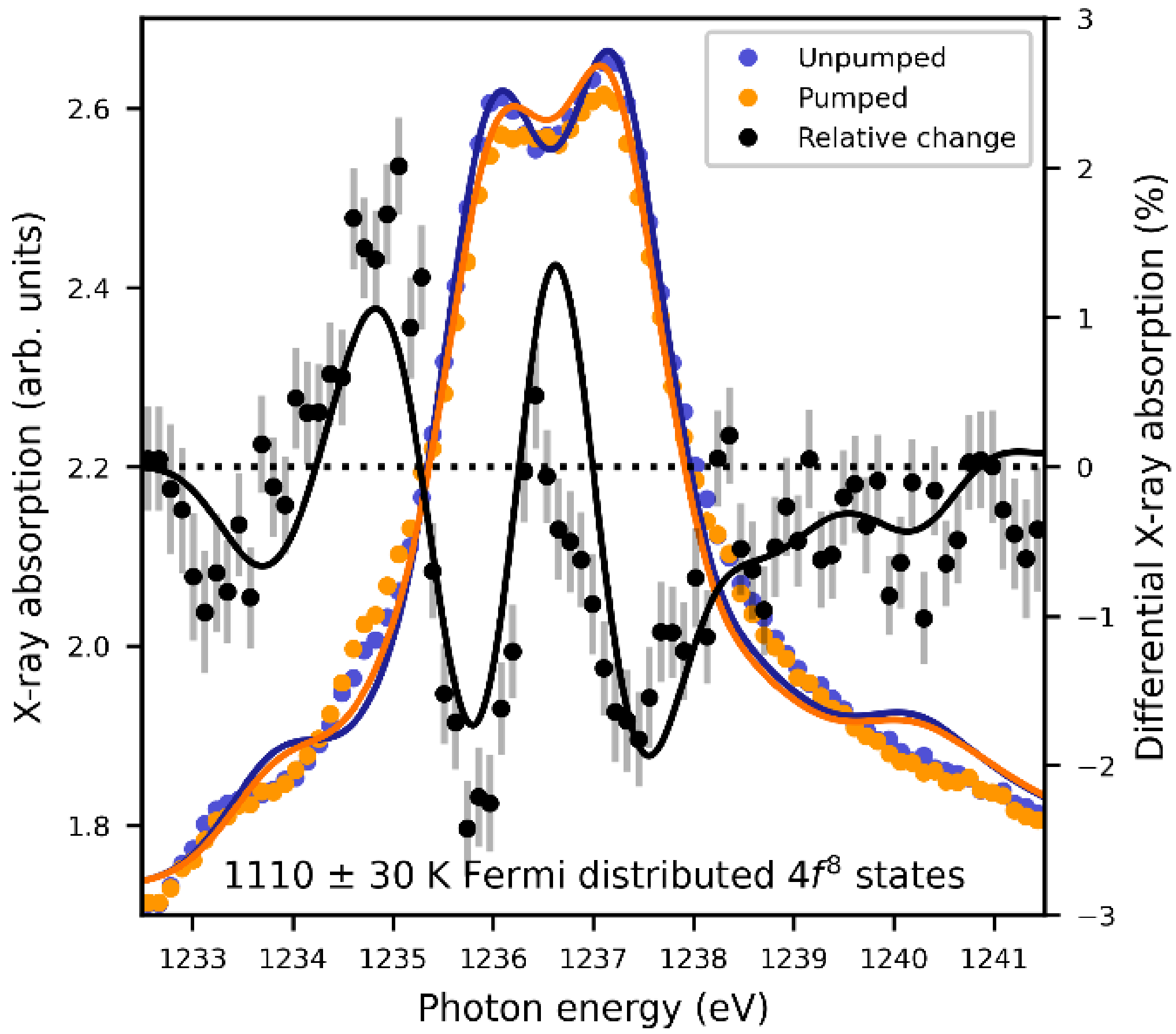

**Fig. S9.**
Tb $M_5$ resonance at 150 fs pump-probe delay, as well as the differential X-ray absorption (dots). The lines show the simulation of the data, based on a fit of the difference signal, assuming admixtures of $4f^8$ excited multiplets to the $^7F_6$ groundstate scaled to the weight of $5d$ electrons in the Fermi distribution that allow for specific $4f^8$ excitations. We find best agreement for a (1110 ± 30) K Fermi profile.

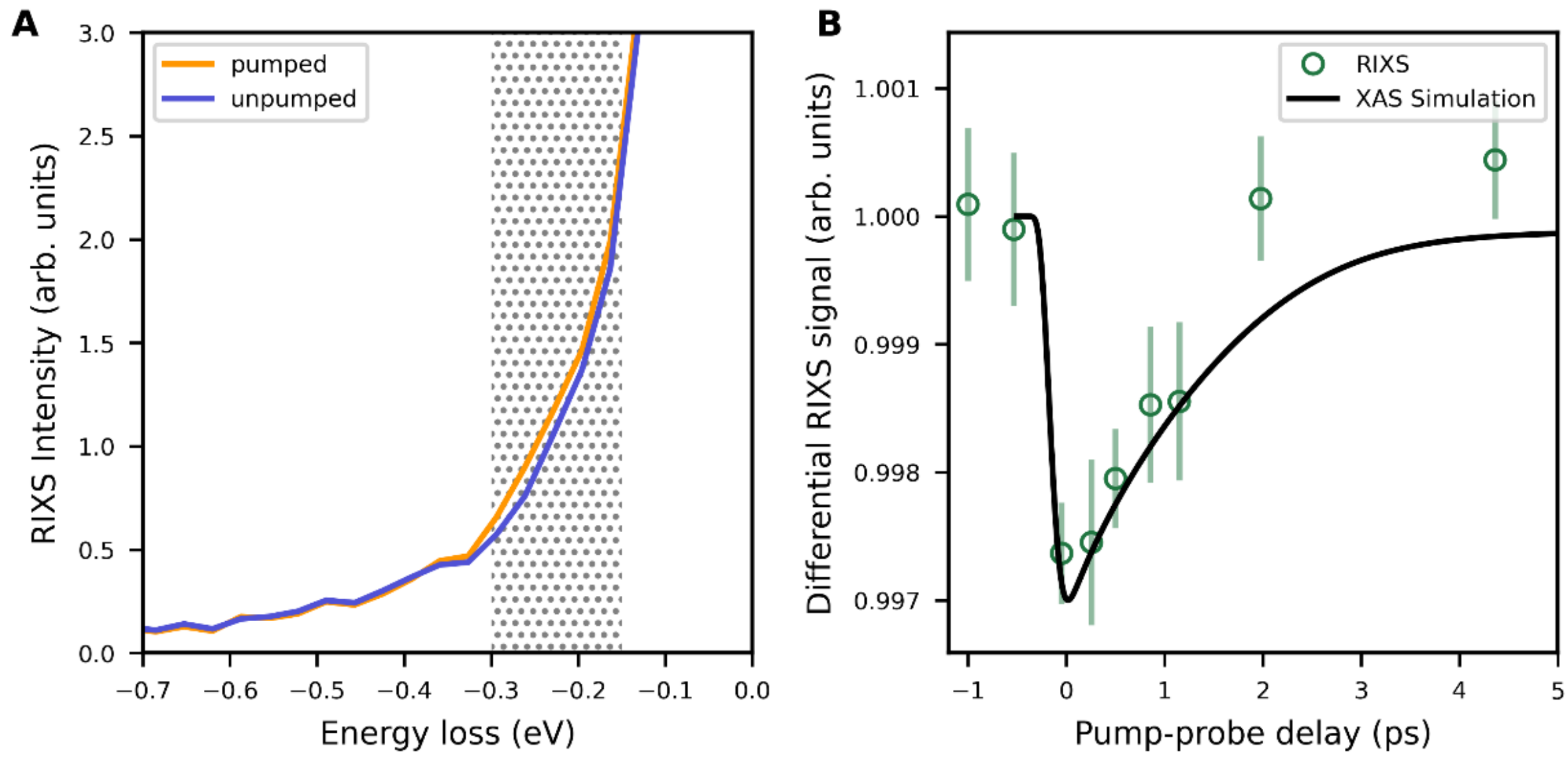

**Fig. S10.**

**(A)** RIXS data for the unpumped (blue line) and pumped (orange line) Tb sample recorded 250 fs after pump-pulse excitation. The incident energy was 149 eV. A weak anti-Stokes feature at the excitation energy of around 260 meV becomes visible ($^7F_6 \rightarrow {}^7F_5$ excitation). **(B)** Pump-probe delay dependence of data integrated over the energy region marked in (A). They follow the transient XAS data, associated with the $^7F_6 \rightarrow {}^7F_5$ excitation.

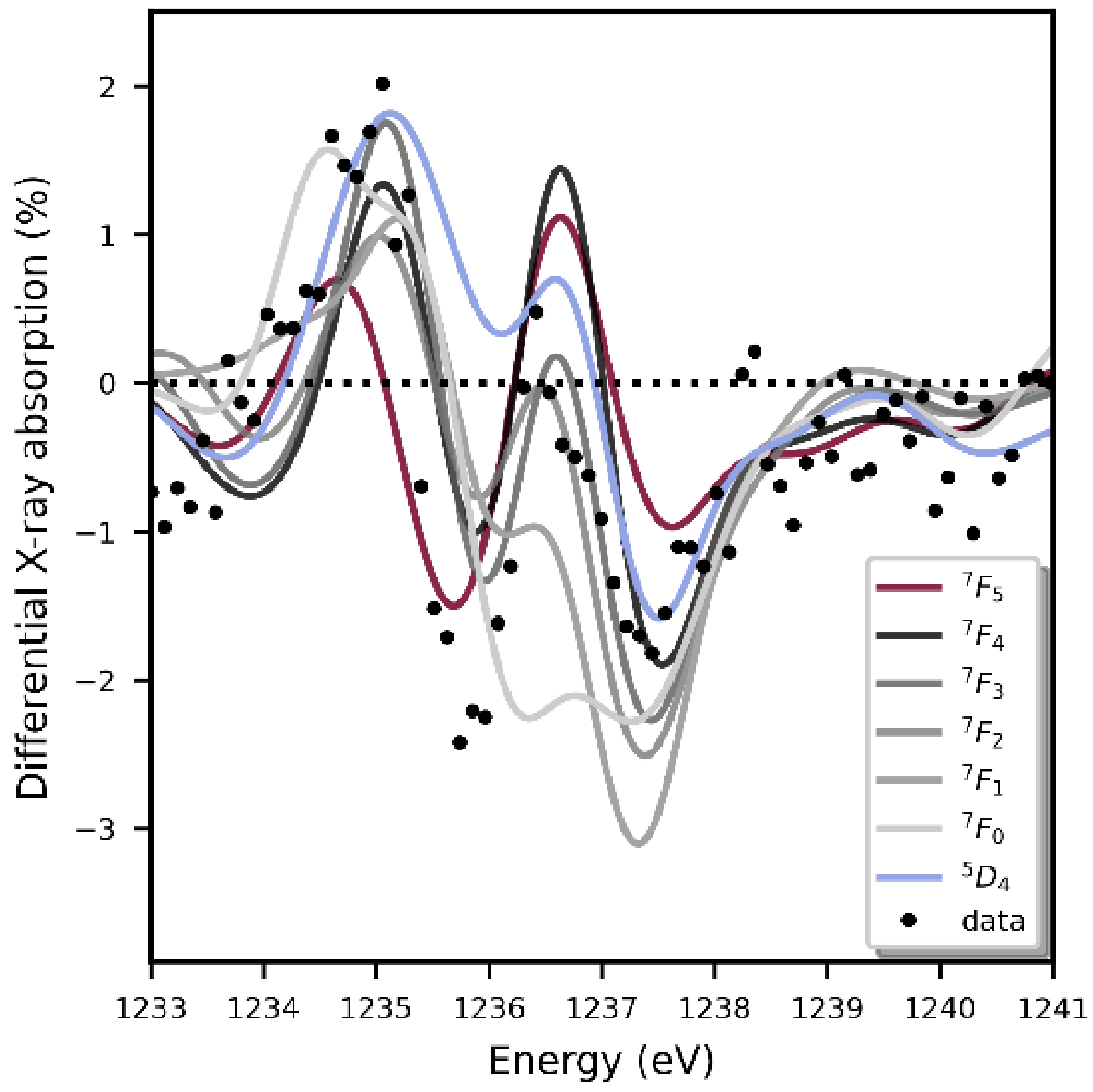

**Fig. S11.**

Differential XAS (black dots) and differential change derived from the multiplet calculation for 15 % admixture of each of the $4f^8$ multiplets (colored lines). Only the $^7F_5$ multiplet exhibits a stronger spectral change at the first main-feature of the $M_5$ resonance (1235.8 eV), the higher lying multiplets show a larger effect at the second peak (1237.5 eV).

| | Tb $4f^7$ | Tb $4f^8$ | Tb $4f^9$ |
|---|---|---|---|
| $f_2$ (ff) | 15.829 | 14.915 | 13.892 |
| $f_4$ (ff) | 9.981 | 9.360 | 8.670 |
| $f_6$ (ff) | 7.195 | 6.734 | 6.225 |
| 4f spin-orbit coupling | 0.237 | 0.221 | 0.205 |

**Table S1.**

Hartree-Fock values (before Slater reduction) for Slater-Condon parameters for different rare-earth electron configurations: $f_i$ parameters are shown for direct 4*f*-4*f* ($f_i$ (*ff*)) interactions (all values in eV).

| | **Tb $3d^9 4f^8$** | **Tb $3d^9 4f^9$** | **Tb $3d^9 4f^{10}$** |
|---|---|---|---|
| $f_2$ (ff) | 16.461 | 15.586 | 14.620 |
| $f_4$ (ff) | 10.390 | 9.794 | 9.141 |
| $f_6$ (ff) | 7.493 | 7.050 | 6.567 |
| 4f spin-orbit coupling | 0.268 | 0.251 | 0.234 |
| 3d spin-orbit coupling | 13.363 | 13.368 | 13.372 |
| $f_2$(df) | 10.631 | 10.055 | 9.468 |
| $f_4$(df) | 5.013 | 4.709 | 4.406 |
| $g_1$(df) | 7.730 | 7.240 | 6.755 |
| $g_3$(df) | 4.535 | 4.245 | 3.959 |
| $g_5$(df) | 3.133 | 2.933 | 2.735 |

## Table S2.

Hartree-Fock Slater-Condon valus for direct 3*d*-4*f* ($f_i$ (*df*)) interactions (all values in eV).

| | Tb $4d^9 4f^8$ | Tb $4d^9 4f^9$ | Tb $4d^9 4f^{10}$ |
|---|---|---|---|
| $f_2$ (ff) | 14.098 | 15.065 | 15.940 |
| $f_4$ (ff) | 8.807 | 9.460 | 10.055 |
| $f_6$ (ff) | 6.326 | 6.808 | 7.249 |
| 4f spin-orbit coupling | 0.208 | 0.249 | 0.241 |
| 4d spin-orbit coupling | 2.366 | 2.386 | 2.413 |
| $f_2$(df) | 16.144 | 16.838 | 17.460 |
| $f_4$(df) | 10.277 | 10.765 | 11.205 |
| $g_1$(df) | 19.012 | 19.842 | 20.588 |
| $g_3$(df) | 11.903 | 12.474 | 12.984 |
| $g_5$(df) | 8.407 | 8.825 | 9.202 |

**Table S3.**

Hartree-Fock Slater-Condon values for direct 4*d*-4*f* ($f_i$ (*df*)) interactions (all values in eV).

| GS | $E_1$ | $E_2$ | $E_3$ | $E_4$ | $E_5$ | $E_6$ | $E_7$ |
| --- | --- | --- | --- | --- | --- | --- | --- |
| $^7F_6$ | $^7F_5$ | $^7F_4$ | $^7F_3$ | $^7F_2$ | $^7F_1$ | $^7F_0$ | $^5D_4$ |
| 0 | 0.28 | 0.41 | 0.54 | 0.62 | 0.68 | 0.71 | 1.71 |

**Table S4.**

Excitation energies ($E_i$ in eV) of the first seven inner shell excitations, for Slater reduction factors of $G_{df} = 0.70$, $F_{df} = 0.80$ and $F_{ff} = 0.61$, as used for the $M_{4,5}$ edge XAS calculations.

| GS | $E_1$ | $E_2$ | $E_3$ | $E_4$ | $E_5$ | $E_6$ | $E_7$ |
|---|---|---|---|---|---|---|---|
| $^7F_6$ | $^7F_5$ | $^7F_4$ | $^7F_3$ | $^7F_2$ | $^7F_1$ | $^7F_0$ | $^5D_4$ |
| 0 | 0.262 | 0.425 | 0.522 | 0.642 | 0.701 | 0.730 | 2.649 |

**Table S5.**

Excitation energies ($E_i$ in eV) oft the first seven inner shell excitations, for Slater reduction factors of $F_{ff}^2 = 0.80$, $F_{ff}^4 = 0.91$ and $F_{ff}^6 = 0.91$, $G_{df} = 0.60$ and $F_{df} = 0.60$, as used for the $N_{4,5}$ edge RIXS calculations.

| $a$ | 1247.26 eV | $d$ | 1.2e-10 m |
|---|---|---|---|
| $\Delta E$ | 0.41 eV | $I_{ej}$ | 0.0641 |
| $E_B$ | 1237.70 eV | $C$ | 1.73 |
| $\Delta E_{lifetime}$ | 0.2 eV | | |

**Table S6.**

Parameters for fitting *Im*[*f(*e)] for the $4f^8\ ^7F_6$ groundstate multiplet to the Tb $M_5$ XAS spectrum for the unpumped sample (see section 7).

```
-- Slater integral reduction may be different for different direct Coulomb and exchange, which can be done by beta1(Fff),
beta2(Fdf) and beta3(Gdf) :
beta1 = 0.61
beta2 = 0.80
beta3 = 0.7
---- Scale the spin-orbit coupling (for direct comparison to CTM4XAS/TT-multiplet 0.99 is chosen)
zeta = 0.99
-- Relative energy axis for the x-ray absorption spectrum (Emin1,Emax1, NE1a amount of points+1 in spectrum); Gamma0
and Lor0 are the Gaussian and Lorentzian broadening parameters for the x-ray absorption spectrum
Emin1 =-30.0
Emax1 = 40.0
NE1a = 7000
Gamma0 = 0.1
Lor0= 0.1
------- Name of output-files for XAS calculations ----
XASfile0="Tb3MXAS_ES00.dat"
XASfile1="Tb3MXAS_ES01.dat"
XASfile2="Tb3MXAS_ES02.dat"
XASfile3="Tb3MXAS_ES03.dat"
XASfile4="Tb3MXAS_ES04.dat"
XASfile5="Tb3MXAS_ES05.dat"
XASfile6="Tb3MXAS_ES06.dat"
XASfile7="Tb3MXAS_ES07.dat"
------- Settings for Tb local interactions -----------------
-- Nelec is the amount of electrons in the 4f shell; F2ff, F4ff, F6ff are the Slater integral parameters for the direct 4f-4f Coulomb
interaction in the initial state; XF2ff, XF4ff and Xf6ff are the Slater integral parameters for the direct 4f-4f Coulomb interaction
in the x-ray excited state (final state of x-ray absorption); F2df,F4df are the Slater integral parameters for the direct 3d-4f
Coulomb interaction (only present in the XAS final state); G1df,G3df are the Slater integral parameters for the 3d-4f exchange
interaction (only present in the XAS final state); zeta_4f and Xzeta_4f are the 4f spin-orbit coupling parameters for the initial
and final state of the x-ray absorption process; zeta_3d is the 3d spin-orbit coupling parameter (which becomes effective in
the final state of the x-ray absorption process, splitting the M4 and M5 edge)
Nelec=8  - - or Nelec=7 - - or Nelec=9
if Nelec==8 then
        F2ff=14.915; F4ff=9.360; F6ff=6.734; zeta_4f=0.221;
        XF2ff=15.586; XF4ff=9.794; XF6ff=7.050; Xzeta_4f=0.251;
        zeta_3d=13.368; F2df=10.055; F4df=4.709; G1df=7.240; G3df=4.245; G5df=2.933;
elseif Nelec==9 then
        F2ff=13.892; F4ff=8.670; F6ff=6.225; zeta_4f=0.205;
        XF2ff=14.620; XF4ff=9.141; XF6ff=6.567; Xzeta_4f=0.234;
        zeta_3d=13.372; F2df=9.468; F4df=4.406; G1df=6.755; G3df=3.959; G5df=2.735;
elseif Nelec==7 then
        F2ff=15.829; F4ff=9.981; F6ff=7.195; zeta_4f=0.237;
        XF2ff=16.461; XF4ff=10.390; XF6ff=7.493; Xzeta_4f=0.268;
        zeta_3d=13.363; F2df=10.631; F4df=5.013; G1df=7.730; G3df=4.535; G5df=3.133;
end
-------- scaling with beta factors (Slater reduction)----------------
-- direct Coulomb ground state (beta1)
F2ff=beta1*F2ff; F4ff=beta1*F4ff; F6ff=beta1*F6ff
-- direct Coulomb p-d excited state (beta2)
F2df=beta2*F2df; F4df=beta2*F4df;
-- exchange p-d (excited state, beta3)
G1df=beta3*G1df; G3df=beta3*G3df; G5df=beta3*G5df
-- direct Coulomb excited state (beta1)
XF2ff=beta1*XF2ff; XF4ff=beta1*XF4ff; XF6ff=beta1*XF6ff
------ scaling with zeta factor (spin-orbit couplings) ----
zeta_3d=zeta*zeta_3d
zeta_4f=zeta*zeta_4f
Xzeta_4f=zeta*Xzeta_4f
------ Number of possible many-body states in the initial configuration (14 f-electrons + 10 d-electrons)
Npsi = math.fact(14) / (math.fact(Nelec) * math.fact(14-Nelec))
NFermion=24
NBoson=0
IndexDn_3d={0,2,4,6,8} -- d-shell [dn]
IndexUp_3d={1,3,5,7,9} -- d-shell [up]
```

```
IndexDn_4f={10,12,14,16,18,20,22} -- f-shell [dn]
IndexUp_4f={11,13,15,17,19,21,23}  -- 4f-shell [up]
--- define operators
OppSx   =NewOperator("Sx"   ,NFermion, IndexUp_4f, IndexDn_4f)
OppSy   =NewOperator("Sy"   ,NFermion, IndexUp_4f, IndexDn_4f)
OppSz   =NewOperator("Sz"   ,NFermion, IndexUp_4f, IndexDn_4f)
OppSsqr =NewOperator("Ssqr" ,NFermion, IndexUp_4f, IndexDn_4f)
OppSplus=NewOperator("Splus",NFermion, IndexUp_4f, IndexDn_4f)
OppSmin =NewOperator("Smin" ,NFermion, IndexUp_4f, IndexDn_4f)
OppLx   =NewOperator("Lx"   ,NFermion, IndexUp_4f, IndexDn_4f)
OppLy   =NewOperator("Ly"   ,NFermion, IndexUp_4f, IndexDn_4f)
OppLz   =NewOperator("Lz"   ,NFermion, IndexUp_4f, IndexDn_4f)
OppLsqr =NewOperator("Lsqr" ,NFermion, IndexUp_4f, IndexDn_4f)
OppLplus=NewOperator("Lplus",NFermion, IndexUp_4f, IndexDn_4f)
OppLmin =NewOperator("Lmin" ,NFermion, IndexUp_4f, IndexDn_4f)
OppJx   =NewOperator("Jx"   ,NFermion, IndexUp_4f, IndexDn_4f)
OppJy   =NewOperator("Jy"   ,NFermion, IndexUp_4f, IndexDn_4f)
OppJz   =NewOperator("Jz"   ,NFermion, IndexUp_4f, IndexDn_4f)
OppJsqr =NewOperator("Jsqr" ,NFermion, IndexUp_4f, IndexDn_4f)
OppJplus=NewOperator("Jplus",NFermion, IndexUp_4f, IndexDn_4f)
OppJmin =NewOperator("Jmin" ,NFermion, IndexUp_4f, IndexDn_4f)
Oppldots=NewOperator("ldots",NFermion, IndexUp_4f, IndexDn_4f)
------ Coulomb operator -----------------
OppF0 =NewOperator("U", NFermion, IndexUp_4f, IndexDn_4f, {1,0,0,0})
OppF2 =NewOperator("U", NFermion, IndexUp_4f, IndexDn_4f, {0,1,0,0})
OppF4 =NewOperator("U", NFermion, IndexUp_4f, IndexDn_4f, {0,0,1,0})
OppF6 =NewOperator("U", NFermion, IndexUp_4f, IndexDn_4f, {0,0,0,1})
--------- Spin-orbit coupling in 3d-shell (core=c)--------
Oppcldots= NewOperator("ldots", NFermion, IndexUp_3d, IndexDn_3d)
------- Core hole potentials -------
---- direct
OppUdfF0 = NewOperator("U", NFermion, IndexUp_3d, IndexDn_3d, IndexUp_4f, IndexDn_4f, {1,0,0}, {0,0,0})
OppUdfF2 = NewOperator("U", NFermion, IndexUp_3d, IndexDn_3d, IndexUp_4f, IndexDn_4f, {0,1,0}, {0,0,0})
OppUdfF4 = NewOperator("U", NFermion, IndexUp_3d, IndexDn_3d, IndexUp_4f, IndexDn_4f, {0,0,1}, {0,0,0})
---- exchange
OppUdfG1 = NewOperator("U", NFermion, IndexUp_3d, IndexDn_3d, IndexUp_4f, IndexDn_4f, {0,0,0}, {1,0,0})
OppUdfG3 = NewOperator("U", NFermion, IndexUp_3d, IndexDn_3d, IndexUp_4f, IndexDn_4f, {0,0,0}, {0,1,0})
OppUdfG5 = NewOperator("U", NFermion, IndexUp_3d, IndexDn_3d, IndexUp_4f, IndexDn_4f, {0,0,0}, {0,0,1})
t=math.sqrt(1/2);
-- setting up the transition operators with various polarisations (dipole x: TXASx, dipole y: TXASy, dipole z: TXASz)
Akm = {{1,-1,t},{1, 1,-t}}
TXASx = NewOperator("CF", NFermion, IndexUp_4f, IndexDn_4f, IndexUp_3d, IndexDn_3d, Akm)
Akm = {{1,-1,t*I},{1, 1,t*I}}
TXASy = NewOperator("CF", NFermion, IndexUp_4f, IndexDn_4f, IndexUp_3d, IndexDn_3d, Akm)
Akm = {{1,0,1}}
TXASz = NewOperator("CF", NFermion, IndexUp_4f, IndexDn_4f, IndexUp_3d, IndexDn_3d, Akm)
--- Input parameters for the Hamiltonian -----
U      =  0.000
F0ff    = U+(4/195)*F2ff+(2/143)*F4ff+(100/5577)*F6ff
XF0ff   = U+(4/195)*XF2ff+(2/143)*XF4ff+(100/5577)*XF6ff
Udf     =  0.000
F0df    =  Udf+G1df*(3/70)+G3df*(2/105)+(5/231)*G5df
--- initial state Hamiltonian
Hamiltonian =  F0ff*OppF0 + F2ff*OppF2 + F4ff*OppF4 + F6ff*OppF6 + zeta_4f*Oppldots
-- final state Hamiltonian
XASHamiltonian = XF0ff*OppF0 + XF2ff*OppF2 + XF4ff*OppF4 + XF6ff*OppF6 +
Xzeta_4f*Oppldots+F0df*OppUdfF0+F2df*OppUdfF2+F4df*OppUdfF4+G1df*OppUdfG1+G3df*OppUdfG3+G5df*OppUdfG
5+zeta_3d*Oppcldots
StartRestrictions = {NFermion, NBoson, {"1111111111 00000000000000",10,10}, {"0000000000
11111111111111",Nelec,Nelec}}
-- Finding the initial state ground state and its optically excited states
psiList = Eigensystem(Hamiltonian, StartRestrictions, Npsi)
for key,value in pairs(psiList) do
  stateenergy = value * Hamiltonian * value
```

```
  print(key, stateenergy)
end
--XAS spectra calculations: Most initial states have degeneracy, so XAS spectra are calculated in a loop over the degenerate
states. The loops for j=x,xx shown below are the examples for Tb3(4f_8). The start and end value of these loops depend on
the 4f_n occupation
SpectraAv = 0
for j=1, 13 do
        TempSpect = CreateSpectra(XASHamiltonian, {TXASx, TXASy, TXASz}, psiList[j], {{"Emin",Emin1}, {"Emax",Emax1},
{"NE",NE1a}})
        SpectraAv = SpectraAv + Spectra.Sum(TempSpect, {1,1,1})/3.
end
Broaden1=Spectra.Broaden(SpectraAv, Gamma0, {{Emin1, Lor0}, {Emax1, Lor0}})
Broaden1.Print({{"file",XASfile0}})
SpectraAv = 0
for j=14, 24 do
        TempSpect = CreateSpectra(XASHamiltonian, {TXASx, TXASy, TXASz}, psiList[j], {{"Emin",Emin1}, {"Emax",Emax1},
{"NE",NE1a}})
        SpectraAv = SpectraAv + Spectra.Sum(TempSpect, {1,1,1})/3.
end
Broaden1=Spectra.Broaden(SpectraAv, Gamma0, {{Emin1, Lor0}, {Emax1, Lor0}})
Broaden1.Print({{"file",XASfile1}})
SpectraAv2 = 0
for j=25, 33 do
        TempSpect = CreateSpectra(XASHamiltonian, {TXASx, TXASy, TXASz}, psiList[j], {{"Emin",Emin1}, {"Emax",Emax1},
{"NE",NE1a}})
        SpectraAv2 = SpectraAv2 + Spectra.Sum(TempSpect, {1,1,1})/3.
end
Broaden2=Spectra.Broaden(SpectraAv2, Gamma0, {{Emin1, Lor0}, {Emax1, Lor0}})
Broaden2.Print({{"file",XASfile2}})
SpectraAv3 = 0
for j=34, 40 do
        TempSpect = CreateSpectra(XASHamiltonian, {TXASx, TXASy, TXASz}, psiList[j], {{"Emin",Emin1}, {"Emax",Emax1},
{"NE",NE1a}})
        SpectraAv3 = SpectraAv3 + Spectra.Sum(TempSpect, {1,1,1})/3.
end
Broaden3=Spectra.Broaden(SpectraAv3, Gamma0, {{Emin1, Lor0}, {Emax1, Lor0}})
Broaden3.Print({{"file",XASfile3}})
SpectraAv4 = 0
for j=41, 45 do
        TempSpect = CreateSpectra(XASHamiltonian, {TXASx, TXASy, TXASz}, psiList[j], {{"Emin",Emin1}, {"Emax",Emax1},
{"NE",NE1a}})
        SpectraAv4 = SpectraAv4 + Spectra.Sum(TempSpect, {1,1,1})/3.
end
Broaden4=Spectra.Broaden(SpectraAv4, Gamma0, {{Emin1, Lor0}, {Emax1, Lor0}})
Broaden4.Print({{"file",XASfile4}})
SpectraAv5 = 0
for j=46, 48 do
        TempSpect = CreateSpectra(XASHamiltonian, {TXASx, TXASy, TXASz}, psiList[j], {{"Emin",Emin1}, {"Emax",Emax1},
{"NE",NE1a}})
        SpectraAv5 = SpectraAv5 + Spectra.Sum(TempSpect, {1,1,1})/3.
end
Broaden5=Spectra.Broaden(SpectraAv5, Gamma0, {{Emin1, Lor0}, {Emax1, Lor0}})
Broaden5.Print({{"file",XASfile5}})
SpectraAv6 = 0
for j=49, 49 do
        TempSpect = CreateSpectra(XASHamiltonian, {TXASx, TXASy, TXASz}, psiList[j], {{"Emin",Emin1}, {"Emax",Emax1},
{"NE",NE1a}})
        SpectraAv6 = SpectraAv6 + Spectra.Sum(TempSpect, {1,1,1})/3.
end
Broaden6=Spectra.Broaden(SpectraAv6, Gamma0, {{Emin1, Lor0}, {Emax1, Lor0}})
Broaden6.Print({{"file",XASfile6}})
SpectraAv7 = 0
for j=50, 58 do
```

```
	TempSpect = CreateSpectra(XASHamiltonian, {TXASx, TXASy, TXASz}, psiList[j], {{"Emin",Emin1}, {"Emax",Emax1},
{"NE",NE1a}})
	SpectraAv7 = SpectraAv7 + Spectra.Sum(TempSpect, {1,1,1})/3.
end
Broaden7=Spectra.Broaden(SpectraAv7, Gamma0, {{Emin1, Lor0}, {Emax1, Lor0}})

Broaden7.Print({{"file",XASfile7}})
```

**Table S7.**

Input file for Quanty calculations (lines starting with "- -" are comment lines).